\pdfoutput=1
\documentclass[journal]{IEEEtran}
\usepackage[utf8]{inputenc}
\usepackage{amsmath}
\usepackage{amsfonts}
\usepackage{amssymb}
\usepackage{graphicx}
\usepackage{siunitx}
\usepackage[hidelinks]{hyperref}

\DeclareGraphicsExtensions{.png,.PNG,.pdf,.PDF}

\DeclareMathOperator{\arctantwo}{arctan2}

\DeclareSIUnit[]{\bit}{bit}

\usepackage[acronym]{glossaries}
\newacronym{psd}{PSD}{power spectral density}
\newacronym{rf}{RF}{radio frequency}
\newacronym{lo}{LO}{local oscillator}
\newacronym{dac}{DAC}{digital-to-analog converter}
\newacronym{adc}{ADC}{analog-to-digital converter}
\newacronym{mcu}{MCU}{microcontroller unit}
\newacronym{pcb}{PCB}{printed circuit board}
\newacronym{ocxo}{OCXO}{oven-controlled crystal oscillator}
\newacronym{pll}{PLL}{phase-locked loop}
\newacronym{cpu}{CPU}{central processing unit}
\newacronym{dds}{DDS}{direct digital synthesizer}
\newacronym{spi}{SPI}{Serial Peripheral Interface}
\newacronym{dsp}{DSP}{digital signal processing}
\newacronym{uart}{UART}{universal asynchronous receiver-transmitter}
\newacronym{fpu}{FPU}{floating-point unit}
\newacronym{usb}{USB}{Universal Serial Bus}
\newacronym{iq}{IQ}{in-phase and quadrature}
\newacronym{crc}{CRC}{cyclic redundancy check}
\newacronym{fir}{FIR}{finite impulse response}
\newacronym{iir}{IIR}{infinite impulse response}
\newacronym{pid}{PID}{proportional-integral-derivative controller}
\newacronym{lpf}{LPF}{low-pass filter}
\newacronym{aom}{AOM}{acousto-optic modulator}
\newacronym{fpga}{FPGA}{field-programmable gate array}
\newacronym{sdr}{SDR}{software-defined radio}
\newacronym{dma}{DMA}{direct memory access interface}
\newacronym{enob}{ENOB}{effective number of bits}
\newacronym{dc}{DC}{direct-coupling}

\title{Embedded digital phase noise analyzer \mbox{for optical frequency metrology}}

\author{Simone Donadello$^*$, Elio K. Bertacco, Davide Calonico and Cecilia Clivati\thanks{The authors are with the Quantum Metrology and Nanotechnologies Division, at INRIM -- Istituto Nazionale di Ricerca Metrologica, 10135 Torino, Italy.

$^*$Corresponding author: Simone Donadello.

Published version DOI: \href{https://doi.org/10.1109/TIM.2023.3288255}{10.1109/TIM.2023.3288255}}}

\begin{document} 
	
\maketitle

\begin{abstract}
	Digital signal processing (DSP) is supporting novel in-field applications of optical interferometry, such as in laser ranging and distributed acoustic sensing. While the highest performances are achieved with field-programmable gated arrays (FPGAs), their complexity and cost are often too high for many tasks. Here, we describe an alternative solution for processing optical signals in real-time, based on a dual-core 32-bit microcontroller. We implemented in-phase and quadrature (IQ) demodulation of optical beat-notes resulting from the interference of independent laser sources, phase noise analysis of deployed optical fibers covering intercity distances, and synchronization of remote acquisitions via optical trigger signals. The embedded architecture can efficiently accomplish a large variety of tasks in the context of optical signal processing, being also easily configurable, compact and upgradable. These features make it attractive for applications that require long-term, remotely-operated, and field-deployed instrumentation.
\end{abstract}


\section{Introduction}

\IEEEPARstart{I}{n recent} years, coherent optical technologies are finding a growing number of applications outside the laboratories where they were first developed: optical clocks emerged as promising quantum sensors for probing the gravitational potential \cite{takamotoTest2020,grottiGeodesy2018}; optically coherent frequency combs suitable for in-field operation have been developed \cite{sinclairOperation2014}, with applications to environmental gas-sensing \cite{riekerFrequencycombbased2014}, high-resolution ranging \cite{caldwellTimeprogrammable2022,bergeronFemtosecond2019}, and transportable clock comparison \cite{gozzardUltrastable2022}; coherent interferometry of narrow-linewidth lasers on deployed optical cables proved to be a powerful tool for the remote distributed sensing of geophysical signals, opening new possibilities for the study of Earth dynamics \cite{marraUltrastable2018,marraOptical2022}, with prospects of a future integration on data transmission networks \cite{ipVibration2022,zhanOptical2021}; finally, highly coherent laser sources in trusted communication nodes could improve the performances of advanced, quantum-secure communication protocols \cite{clivatiCoherent2022}.

Many of these applications require high-resolution and coherent measurements of the optical signal phase. While this is an easy task in a laboratory equipped with commercial instruments like frequency counters or phase noise analyzers, the development of portable, reconfigurable, and scalable measurement setups equipped with remote control and alert systems is critical for in-field applications. Digital electronics and \gls{sdr} greatly support this challenge, fulfilling sophisticated and time-critical tasks, and allowing preliminary signal processing onboard, before relevant data are transferred to remote control units at suitably reduced rates. So far, custom platforms based on \glspl{fpga} emerged as the best solutions for these tasks, with many applications in the context of optical frequency metrology \cite{shermanOscillator2016,mukherjeeDigital2022,clivatiRobust2020}. Nevertheless, their high cost and the complexity associated to the board configuration makes them a not obvious nor optimal choice in many cases, especially when compared to more affordable and practical architectures \cite{tourigny-planteOpen2018,yangLowcost2012,elaskarFPGABased2022}.

Here we describe another approach, that makes use of a \gls{mcu} to implement optical phase measurements, based on the heterodyne demodulation scheme inherited from digital lock-in amplifiers \cite{baroneHigh1995,kishoreEvolution2020,zhangFPGABased2020}. The platform is based on the affordable yet high-performance STM32H7 unit, with dual-core 32-bit ARM architecture and integrated \glspl{adc}. Custom electronics and firmware were developed to realize a compact low-cost phase noise analyzer, with reconfigurable bandwidth and sampling rate. The optimized algorithm implements \gls{iq} phase detection. Carrier frequency and amplitude of \gls{rf} signals can be measured with high-resolution, at output data rates up to \SI{20}{\kilo\hertz}. The system is clocked with a stable quartz reference, and it includes synthesizers, \glspl{dac} and \glspl{dds} for signal conditioning, resulting in a fully-embedded stand-alone remote measurement unit.

Important features have been developed for the realization of long-term experiments with real-time acquisitions. In particular, we focused on developing a platform suitable for the frequency comparison of laser sources with spectral separation up to a few gigahertz, possibly varying over time, and the analysis of phase noise accumulated by a coherent signal as it travels a long-haul fiber. We implemented routines for synchronizing remote acquisitions within microseconds, suppressing common-mode noise processes. Our \gls{dsp} solution addresses the typical challenges encountered in the processing of coherent optical signals, with proper handling of the amount of phase noise entering the demodulation steps, and combining high data rates with flexible and lightweight firmware algorithms.

\section{System architecture}

\subsection{Electronics and MCU}

\begin{figure*}[!t]
	\centering
	\includegraphics[width=1.66\columnwidth]{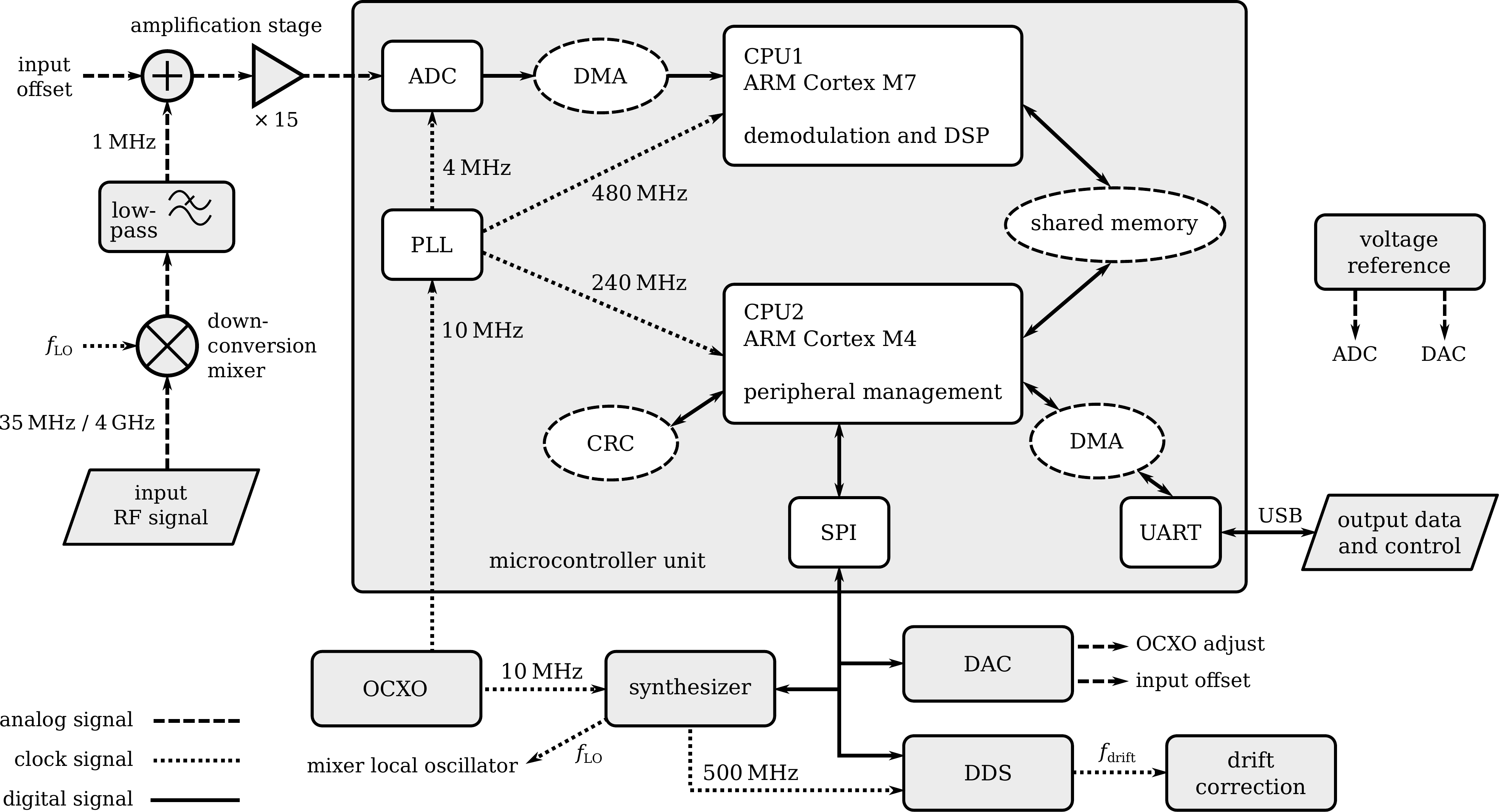}
	\caption{Scheme of the acquisition system based on a dual-core microcontroller and its peripherals; the main elements of the input signal conditioning stage and other onboard components are also reported.}
	\label{fig:scheme-micro}
\end{figure*}

Our digital phase analyzer is implemented on a dual-core \gls{mcu} from the STM32 series (STM32H745, STMicroelectronics), based on ARM Cortex \glspl{cpu} with M7 and M4 32-bit architectures (CPU1 and CPU2 respectively). A custom \gls{pcb} has been designed as a shield for the \gls{mcu} development board NUCLEO-H745ZI-Q. The \gls{pcb} includes the conditioning electronics for the input analog signals, as well as all the components required for the embedded system functioning. The electronic scheme of the microcontroller with its internal and external peripherals is sketched in Fig. \ref{fig:scheme-micro}.

The \gls{mcu} includes $3$ multi-channel \glspl{adc}, having 16-bit nominal resolution and \num{12.2} \gls{enob}. The input signal is acquired using a fast \gls{adc} channel operated at 14-bit, which allows a maximum sampling rate of \SI{5}{\mega\hertz}. To match the input bandwidth requirements, the signal to be measured is down-converted to the nominal frequency $\nu_0=\SI{1}{\mega\hertz}$ using an external frequency mixer in combination with a \gls{lpf} with \SI{1.9}{\mega\hertz} cutoff. The \gls{rf} signal is amplified and brought to the \gls{adc} voltage range of \SIrange{0}{2.5}{\volt} using a bipolar operational amplifier in non-inverting configuration (AD8027, Analog Devices), with fixed gain $15$, bandwidth from \SIrange{0}{8}{\mega\hertz}, and configurable voltage offset. This \gls{adc} input amplification scheme allows to acquire signals with different polarizations and frequencies, useful for debugging and general-purpose applications. For instance, when the external mixer is removed, this scheme also supports low-noise \gls{dc} measurements.

The \gls{lo} signal for the frequency mixer is either generated by an embedded synthesizer (ADF4351, Analog Devices) in the range from \SI{35}{\mega\hertz} to \SI{4.4}{\giga\hertz}, or by a \gls{dds} (AD9912, Analog Devices) whose frequency is programmed with 48-bit resolution by the \gls{mcu} via \gls{spi}. This ensures on one side a broadband operating range for the input signal, and on the other the possibility of slowly tracking its low-frequency drifts by finely adjusting the \gls{dds} tuning word.

A \SI{10}{\mega\hertz} \gls{ocxo} is integrated on the custom shield as the master clock source for the \gls{mcu} and the other components. A good clock source is crucial for a high-performance phase measurement system. The chosen \gls{ocxo} (HCD660/FTFN, Golledge) is characterized by \num{2E-10} aging per day, short term stability better than \num{1e-12} (Allan deviation at \SI{1}{\second}), and phase noise lower than \SI{-155}{dBc\per\hertz} above \SI{1}{\kilo\hertz}. The \gls{mcu} clock signals are generated from the \SI{10}{\mega\hertz} reference using the internal \glspl{pll}. These include the \SI{480}{\mega\hertz} clock of CPU1 and the \SI{240}{\mega\hertz} clock for CPU2. A synchronous \SI{4}{\mega\hertz} clock is generated by a \gls{mcu} hardware timer, and used to trigger the \gls{adc} acquisition. The synthesizer at the input down-conversion stage is also referenced to the same \gls{ocxo}, while the \gls{dds} is clocked at \SI{500}{\mega\hertz} by the synthesizer. An external \SI{10}{\mega\hertz} source can be provided for applications demanding a common clock architecture or higher accuracy, replacing the \gls{ocxo} in all the above-mentioned tasks.

A 16-bit \gls{dac} (AD5686, Analog Devices) is used to generate the reconfigurable voltage offset for the signal conditioning stage, and to fine-adjust the \gls{ocxo} frequency through its dedicated tuning voltage input. The latter is used to compensate at the sub-\si{\milli\hertz} level the \gls{ocxo} frequency offset and the quartz aging relatively to a reference oscillator.

The board is powered with \SI{\pm 15}{\volt} and cooled with forced air flow. A low-noise precision voltage source at $V_\text{ref}=\SI{2.5}{\volt}$ (MAX6225-AE, Maxim Integrated) is used as analog reference for the \gls{dac} and the \gls{mcu} internal \glspl{adc}. Other sensors can be connected to the \gls{mcu} shield by means of \gls{spi}, such as accelerometers and temperature sensors used to monitor the measurement environment conditions. Additional \gls{adc} channels can be used to acquire other experimental parameters at lower frequencies.

The \gls{mcu} firmware is programmed in C language, exploiting the STM32 HAL and Low-Layer libraries. The dual-core architecture of the microcontroller is organized with the high-speed CPU1 being used for the computationally-demanding \gls{dsp} operations to be performed in real-time. These include \gls{adc} readout, \gls{iq} demodulation, digital filtering, and decimation to lower sampling rates. The computational cost is reduced by taking advantage of the \gls{dma} for data transfer between the internal peripherals, which can operate without consuming \gls{cpu} resources. Moreover, the data and instruction cache embedded in the \gls{mcu}, in conjunction with a careful optimization of the firmware loops, reduces the memory access time while processing data. Finally the 32-bit \gls{fpu} allows faster calculations, with many floating-point operations being performed as single \gls{cpu} cycles.

After decimation, the processed data are transferred to CPU2 exploiting a shared memory region and hardware semaphores. That secondary processor is dedicated to the continuous data streaming toward a computer unit, used for data storage. The connection is obtained with a \gls{uart} serial interface via \gls{usb} at \SI{4}{\mega\bit\per\second} baud rate. Data consistency is checked by means of the \gls{crc} internal peripheral. CPU2 is dedicated also to the general system management: it parses the commands received from the computer to configure and control the acquisition, and it manages the external peripherals connected to the board via \gls{spi}. Operation priorities are managed by means of \gls{mcu} hardware interrupts.

\subsection{Digital signal processing}

\subsubsection{Phase demodulation}

The input signal phase is processed exploiting the \gls{iq} demodulation technique. This is a common approach to phase detection in digital systems, typically implemented on \glspl{fpga} to process the \gls{rf} signals acquired with fast \glspl{adc}  \cite{gervasoniSwitched2017,stimpsonOpensource2019,wangHighPerformance2022}. However, some algorithm optimizations have to be considered to maintain good performances also on a resource-limited \gls{mcu} such as in our case  \cite{nooruddinSimple2020,dorringtonSimple2002,leisSimplified2012,zhangOptimization2016}.

It is useful to first recap a few general concepts that will help to illustrate our \gls{dsp} operation. First, consider the general case of a \gls{rf} voltage signal $V(t)$ at time $t$, characterized by nominal frequency $\nu_0$, and time-varying amplitude $A_\text{s}(t)$ and phase~$\phi(t)$, expressed as
\begin{equation}
	V(t) = A_\text{s}(t) \sin(2 \pi \nu_0 t + \phi(t) )\,.
\end{equation}
Time-dependent fluctuations of the signal phase give rise to deviations of the instantaneous carrier frequency $\nu(t)$ from the nominal value $\nu_0$, and are the primary information of interest in our context, thus
\begin{equation}
	\label{eq:phase-variation}
	\Delta \nu(t) = \nu(t) - \nu_0 = \frac{1}{2\pi}\frac{d\phi(t)}{dt}\,.
\end{equation}
The two notations will be indifferently adopted throughout the text.

In a phase-sensitive detector, the input signal is demodulated by mixing to the in-phase and quadrature references $r_\text{I}(t)$ and $r_\text{Q}(t)$, here conveniently taken at frequency $\nu_0$ and with unitary amplitude
\begin{subequations}
\begin{align}
	r_\text{I}(t) &= \cos(2\pi \nu_0 t)\\
	r_\text{Q}(t) &= \sin(2\pi \nu_0 t)\,.
\end{align}
\end{subequations}
The mixed signals are low-pass filtered with transfer function $H_\text{LPF}$ and cutoff frequency $f_\text{BW}$, to extract the \gls{iq} components:
\begin{subequations}
\label{eq:iq-analog}
\begin{align}
	I(t) &= H_\text{LPF}\{r_\text{I}(t) V(t)\} = \frac{1}{2}A_\text{s}(t)\cos(\phi(t))\\
	Q(t) &= H_\text{LPF}\{r_\text{Q}(t) V(t)\} = \frac{1}{2}A_\text{s}(t)\sin(\phi(t))\,.
\end{align}
\end{subequations}
The filter suppresses the higher-order frequencies, and determines the demodulation bandwidth. The signal amplitude and phase are then calculated as
\begin{subequations}
\label{eq:ampl-phase-analog}
\begin{align}
	A_\text{s}(t) &= 2\sqrt{I^2(t)+Q^2(t)}\\
	\phi'(t) &= \arctantwo(Q(t), I(t))\,.
\end{align}
\end{subequations}
It is worth to notice that $\phi'(t)$ is a wrapped version of the original phase $\phi(t)$, since the domain of the four-quadrant inverse tangent function is defined between $-\pi$ and $+\pi$.

\begin{figure*}[!t]
	\centering
	\includegraphics[width=2\columnwidth]{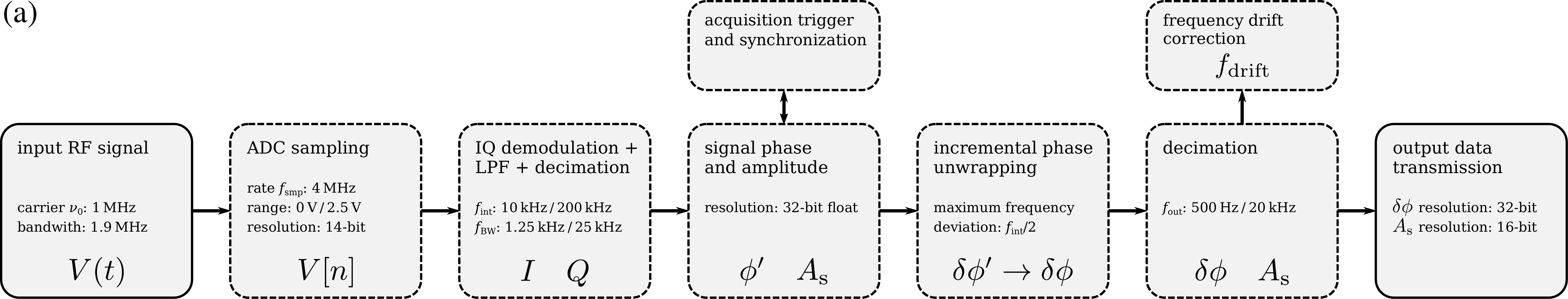}\\[0.5em]
	\includegraphics[width=1\columnwidth]{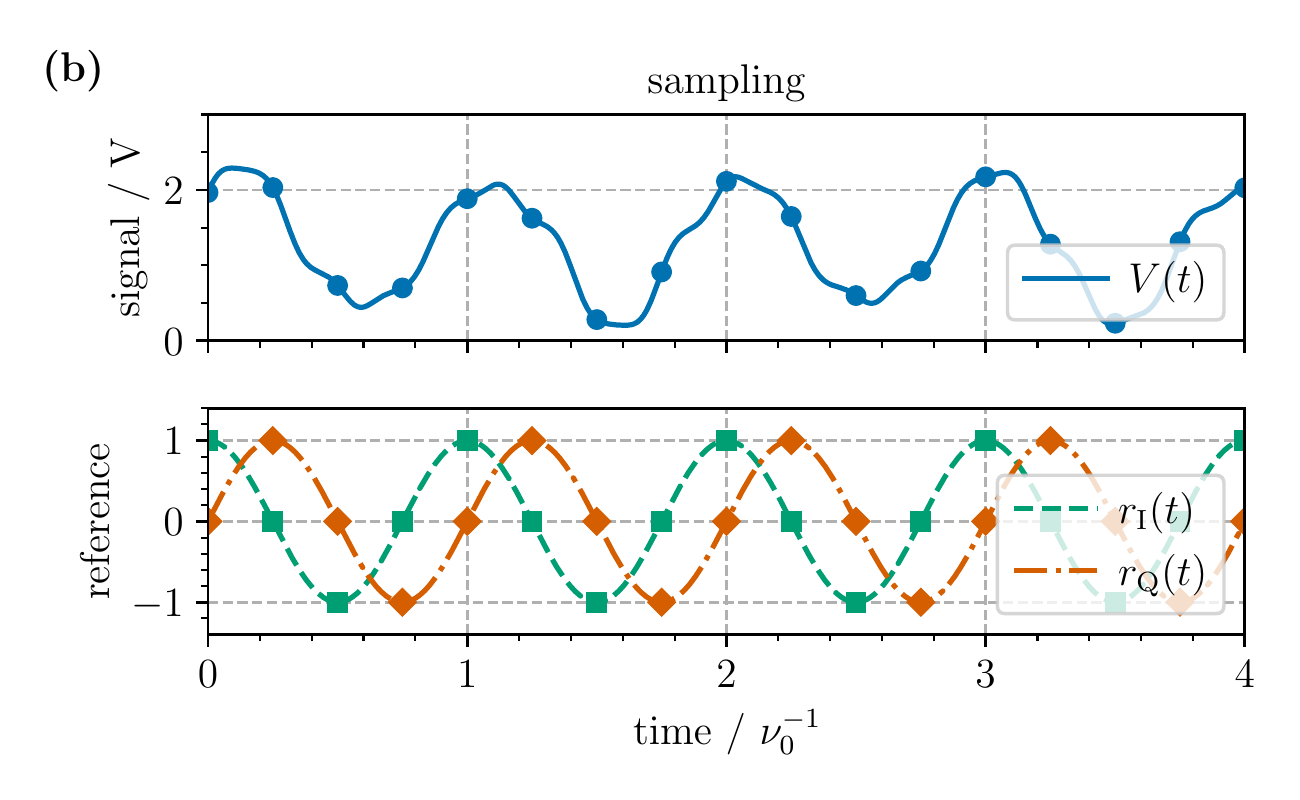}\hfill
	\includegraphics[width=1\columnwidth]{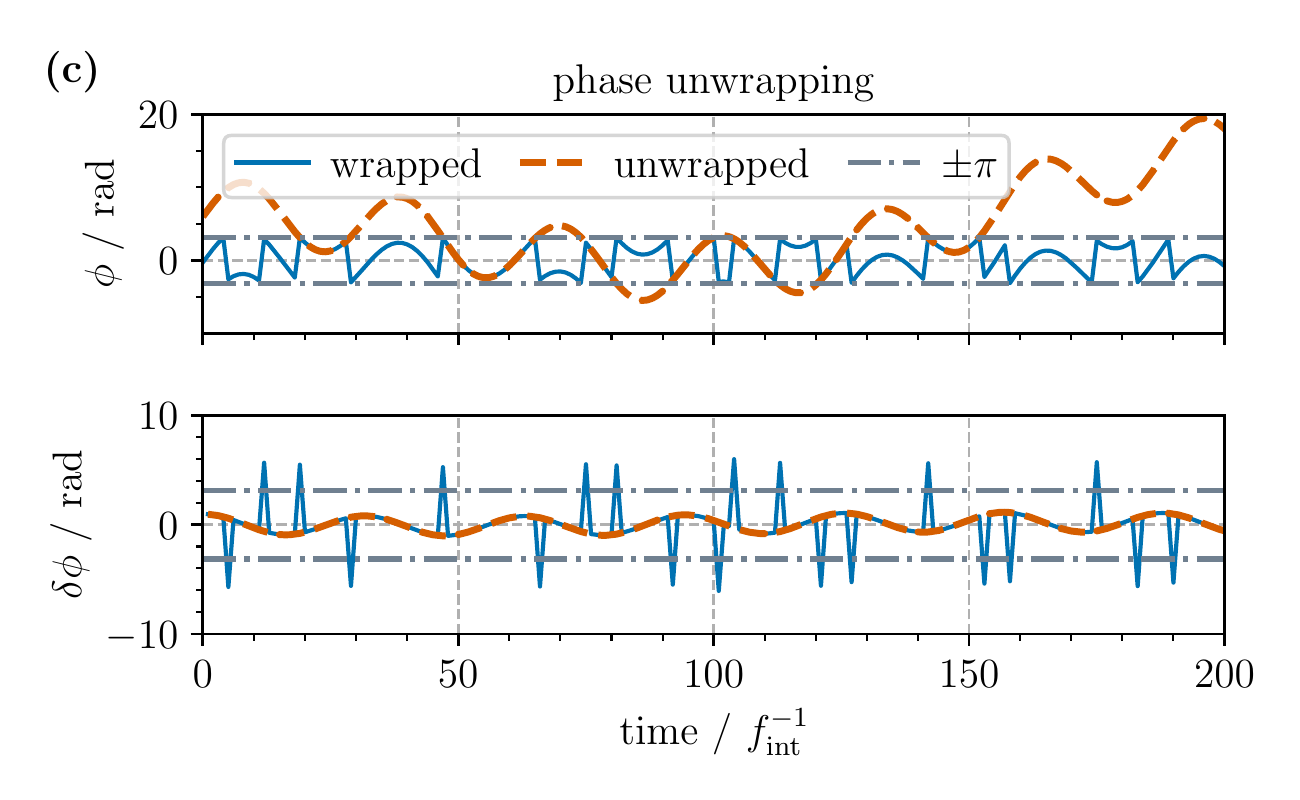}
	\caption{(a) Scheme of the main \glsentryshort{dsp} steps implemented in the microcontroller firmware for signal demodulation. (b) Sketch of the sampled input signal, demodulated relatively to internal \glsentryshort{iq} references. (c) Sketch of phase unwrapping to reconstruct the original absolute phase.}
	\label{fig:scheme-dsp}
\end{figure*}

These same steps are implemented in our \gls{mcu} firmware, as shown in Fig. \ref{fig:scheme-dsp}(a). When the signal $V(t)$ is sampled by the \gls{adc} at frequency $f_\text{smp}$, the previous relations remain valid at discrete instants $n / f_\text{smp}$, with $n$ integer, thus
\begin{equation}
	V[n] = A_\text{s}[n] \sin(2 \pi \nu_0 n / f_\text{smp} + \phi[n])\,.
\end{equation}
The \gls{iq} references are generated numerically, hence driven by the \gls{ocxo} clock and the \gls{mcu} \glspl{pll}, such that
\begin{subequations}
\label{eq:demod-taps}
\begin{align}
	r_\text{I}[n] &= \cos(2\pi \nu_0 n / f_\text{smp})\\
	r_\text{Q}[n] &= \sin(2\pi \nu_0 n / f_\text{smp})\,.
\end{align}
\end{subequations}

The \gls{lpf} with cutoff $f_\text{BW}$ has been implemented as a \gls{fir} filter. Although its higher computational cost, only partially mitigated by the \gls{mcu} cache and \gls{fpu}, the choice of a \gls{fir} filter over a \gls{iir} one is mainly justified by its linear phase response: this is a key feature for accurate phase measurements without distortion. Moreover, due to absence of feedback, the intrinsic stability of \gls{fir} filters makes them the optimal solution for reliable acquisitions, avoiding effects of self-oscillations and coefficient quantization errors.

With a \gls{fir} implementation, the filtered signal becomes a weighted sum of $N+1$ terms, with $N$ the filter order, and the weight coefficients defining the frequency response. Consequently, the algorithm complexity can be reduced by exploiting the system linearity, specifically by combining the \gls{lpf} transfer function $H_\text{LPF}$ with the reference mixing operation in \eqref{eq:iq-analog}. This means that the demodulation of the signal components $I[n]$ and $Q[n]$ can be performed with single \gls{fir} filtering operations, having transfer functions $H_\text{I}$ and $H_\text{Q}$ respectively, such that
\begin{subequations}
\begin{align}
\begin{split}
	I[n] &= H_\text{I}\{V[n]\} \\&= \sum_{i=0}^{N}h_\text{I}[i]V[n-i] = \frac{1}{2}A_\text{s}[n]\cos(\phi[n])
\end{split}\\
\begin{split}
	Q[n] &= H_\text{Q}\{V[n]\} \\&= \sum_{i=0}^{N}h_\text{Q}[i]V[n-i] = \frac{1}{2}A_\text{s}[n]\sin(\phi[n])\,.
\end{split}
\end{align}
\end{subequations}
Exploiting symmetry properties, the demodulation filter coefficients $h_\text{I}[i]$ and $h_\text{Q}[i]$ are calculated as the product between the \gls{lpf} coefficients $h_\text{LPF}[i]$ and the references $r_\text{I}[i]$ and $r_\text{Q}[i]$, respectively:
\begin{subequations}
	\begin{align}
		h_\text{I}[i] &= r_\text{I}[i]h_\text{LPF}[i]\\
		h_\text{Q}[i] &= r_\text{Q}[i]h_\text{LPF}[i]\,.
	\end{align}
\end{subequations}

It is worth to notice that, although sharing the same \gls{lpf} coefficients and frequency response, characterized by bandwidth $f_\text{BW}$ and center frequency $\nu_0$, the combination of filtering with \gls{iq} reference mixing allows to discriminate the two orthogonal signal components, shifted by $f_\text{smp} / 4 \nu_0$ samples. Compared to independent mixing and filtering, this demodulation approach requires fewer real-time computations for each sample.

Considering that the \gls{iq} components are limited to a bandwidth $f_\text{BW}$, they can be decimated without aliasing down to a reduced rate $f_\text{int}>2f_\text{BW}$. Therefore only one sample every $f_\text{smp}/f_\text{int}$ needs to be processed, effectively reducing the data rate and \gls{cpu} load. Amplitude $A_\text{s}[n]$ and wrapped phase $\phi'[n]$ are then calculated at the intermediate rate $f_\text{int}$ using \eqref{eq:ampl-phase-analog}.

The algorithm is further optimized by choosing
\begin{equation}
	\label{eq:freq-sampling}
	f_\text{smp}=4 \nu_0\,.
\end{equation}
In fact, as it can be seen from \eqref{eq:demod-taps} and Fig. \ref{fig:scheme-dsp}(b), in this case half of the \gls{iq} coefficients are equal to zero. This means that the actual computational cost is halved, or that the filter order of $H_\text{I}$ and $H_\text{Q}$ can be doubled for a given \gls{cpu} clock frequency.

\subsubsection{Phase unwrapping}

Following the $\arctantwo$ function definition, $\phi'[n]$ wraps with $\pm 2\pi$ jumps when the signal phase exceeds the $\mp \pi$ domain bounds \cite{itohAnalysis1982}. An unwrapping algorithm is required to recognize and compensate such phase discontinuities, recovering the continuous phase as sketched in Fig. \ref{fig:scheme-dsp}(c).

The processing is performed on the phase increments, calculated from the absolute phase values:
\begin{equation}
	\delta \phi'[n] = \phi'[n]-\phi'[n-1]\,.
\end{equation}
Phase wraps can be detected by searching for values of wrapped phase increments $\delta \phi'[n]$ larger than $\pm\pi$. Consequently, the unwrapped phase increments $\delta \phi[n]$ can be reconstructed:
\begin{equation}
	\begin{cases}
		\delta \phi'[n] > +\pi&: \quad \delta \phi[n] = \delta \phi'[n] - 2\pi\\
		-\pi \leq \delta \phi'[n] \leq +\pi&: \quad \delta \phi[n] = \delta \phi'[n]\\
		\delta \phi'[n] < -\pi&: \quad \delta \phi[n] = \delta \phi'[n] + 2\pi\,.
	\end{cases}
\end{equation}
The absolute phase can be retrieved by integration as
\begin{equation}
	\phi[n] = \sum_{i=1}^{n}\delta \phi[i]\,.
\end{equation}
Nonetheless the samples are always treated in terms of phase increments during the whole \gls{dsp} process, since the absolute phase could accumulate rapidly to big values, and floating-point variables with 32-bit resolution would loose precision soon. On the contrary, phase increments are confined within the $\pm\pi$ interval.

Considering that unwrapping is performed for each new value of $\delta \phi'[n]$ at the intermediate frequency $f_\text{int}$, from \eqref{eq:phase-variation} it follows that the instantaneous frequency deviation from the nominal carrier $\nu_0$ is proportional to the unwrapped incremental phase $\delta \phi[n]$, thus
\begin{equation}
	\Delta \nu[n] = \nu[n] -\nu_0 = \delta \phi[n] \frac{f_\text{int}}{2 \pi}\,.
\end{equation}

In applications where optical signals are affected by phase noise processes, it is critical to preserve the coherence of the accumulated phase \cite{volkovDeterministic2003}. This can be broken by two distinct effects which are addressed in the following. First of all, big instantaneous frequency deviations, giving rise to $|\delta \phi[n]| > \pi$, can be mistakenly interpreted as phase wraps and cause aliasing. The maximum frequency deviation that can be unwrapped without ambiguity is
\begin{equation}
	\Delta \nu_\text{max} = \pm \frac{f_\text{int}}{2}\,.
\end{equation}
Phase undersampling can be avoided by increasing $f_\text{int}$. However the intermediate frequency is typically limited by the computational power of the processing unit. A good demodulation filtering stage which attenuates the high frequency components can mitigate the aliasing effects.

Another aspect that can cause phase unwrap failures is phase noise. In fact, high noise levels added to the original signal will cause wrong unwrap events. A solution would be a narrower demodulation bandwidth to suppress a wider part of the noise spectrum. On the other hand, $f_\text{BW}$ must be chosen depending on the actual signal dynamics: if the demodulation filter band is too narrow, the demodulated signal can be distorted. Moreover, narrower filtering requires higher filter orders: at finite computational power this can be achieved only at lower $f_\text{int}$ values. Therefore a compromise between this and the previous limits must be taken into account.

\subsubsection{Data decimation}

Although phase unwrapping must be performed at high rate to avoid aliasing, the required output data rate and measurement bandwidth are typically much lower than $f_\text{int}$, which can be of the order of several tens of \si{\kilo\hertz}. Therefore, after demodulation and phase unwrapping, amplitude and incremental phase are further decimated to frequency $f_\text{out}$ by applying an anti-aliasing \gls{fir} filter, and by taking one sample every $f_\text{int}/f_\text{out}$. The decimated data are then packed into chunks of samples, and transmitted to the control computer. Time consistency of samples is guaranteed by the \gls{ocxo} clock over all the \gls{dsp} steps.

\subsubsection{Acquisition trigger}

The \gls{mcu} supports external triggering through a digital input assigned to a hardware interrupt that starts the acquisition. Nevertheless some applications require synchronous acquisitions at remote locations, without a direct access to the trigger signal: to meet this requirement, we enabled the system to accept trigger signals encoded as specific modulation patterns onto the acquired signal. A dedicated firmware routine manages the acquisition start: initially the demodulation process is kept running in an idle state, where either amplitude or phase are continuously sampled and compared with the required trigger level at rate $f_\text{int}$; the actual data transmission, hence the acquisition timescale, is started only when the trigger condition is met. Therefore, the acquisition timescales of independent boards measuring a common signal can be synchronized, with a maximum relative delay of $\pm 1/f_\text{int}$ given by the condition check rate.

\subsubsection{Drift correction}
\label{sec:drift}

Our implementation of \gls{iq} demodulation operates with a digital heterodyne reference, fixed by the \gls{ocxo} clock. If the carrier frequency of the input signal differs significantly from the nominal value $\nu_0$ or changes over time, namely by more than a quantity $f_\text{BW}$, it can exceed the demodulation filter band, and the acquisition fails. To avoid this, a digital \gls{pid} is implemented in the \gls{mcu} firmware to track and compensate in real-time the slow carrier drifts. The \gls{pid} takes as input the demodulated frequency deviation passed through an appropriate \gls{lpf}, and it adjusts the \gls{dds} tuning word at frequency $f_\text{drift}$. This can be used as negative feedback to drive either the \gls{lo} of the input down-conversion mixer or, e.g., an external modulator that actually shifts the carrier frequency. The original signal frequency can be reconstructed deterministically in post-processing by combining the measurements with the recorded $f_\text{drift}$ values.

\subsection{System configuration}

\subsubsection{DSP parameters}

The carrier center frequency is designed as $\nu_0=\SI{1}{\mega\hertz}$. This is a compromise between the maximum \gls{adc} sampling rate and the \gls{dsp} capabilities on one hand, and the lower electronic noise affecting the high-frequency spectrum. Higher central frequencies also enable wider dynamic ranges to be correctly demodulated. Following \eqref{eq:freq-sampling}, the \gls{adc} sampling is triggered by an internal hardware timer at $f_\text{smp}=4\nu_0=\SI{4}{\mega\hertz}$.

\begin{figure}[!t]
	\centering
	\includegraphics[width=1\columnwidth]{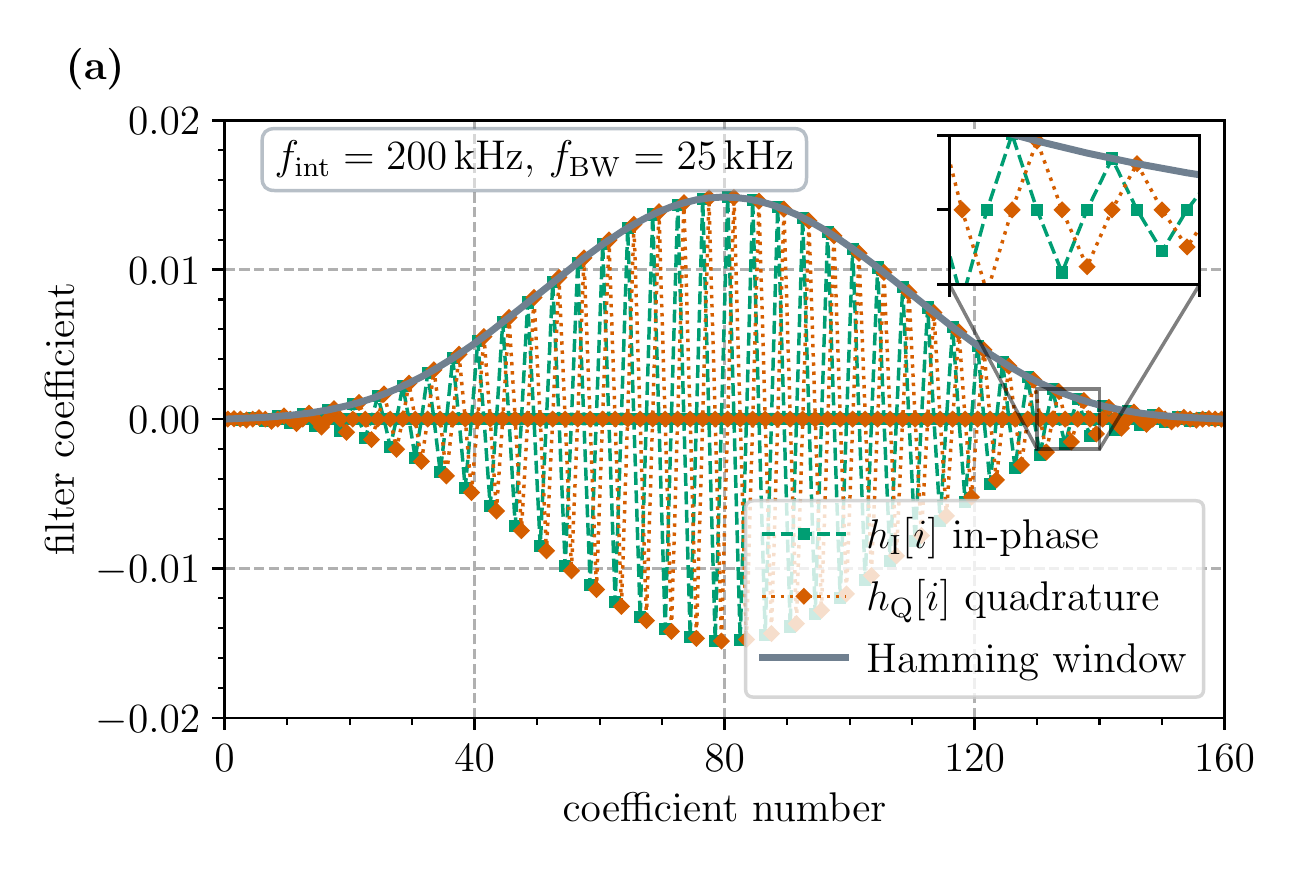}\\
	\includegraphics[width=1\columnwidth]{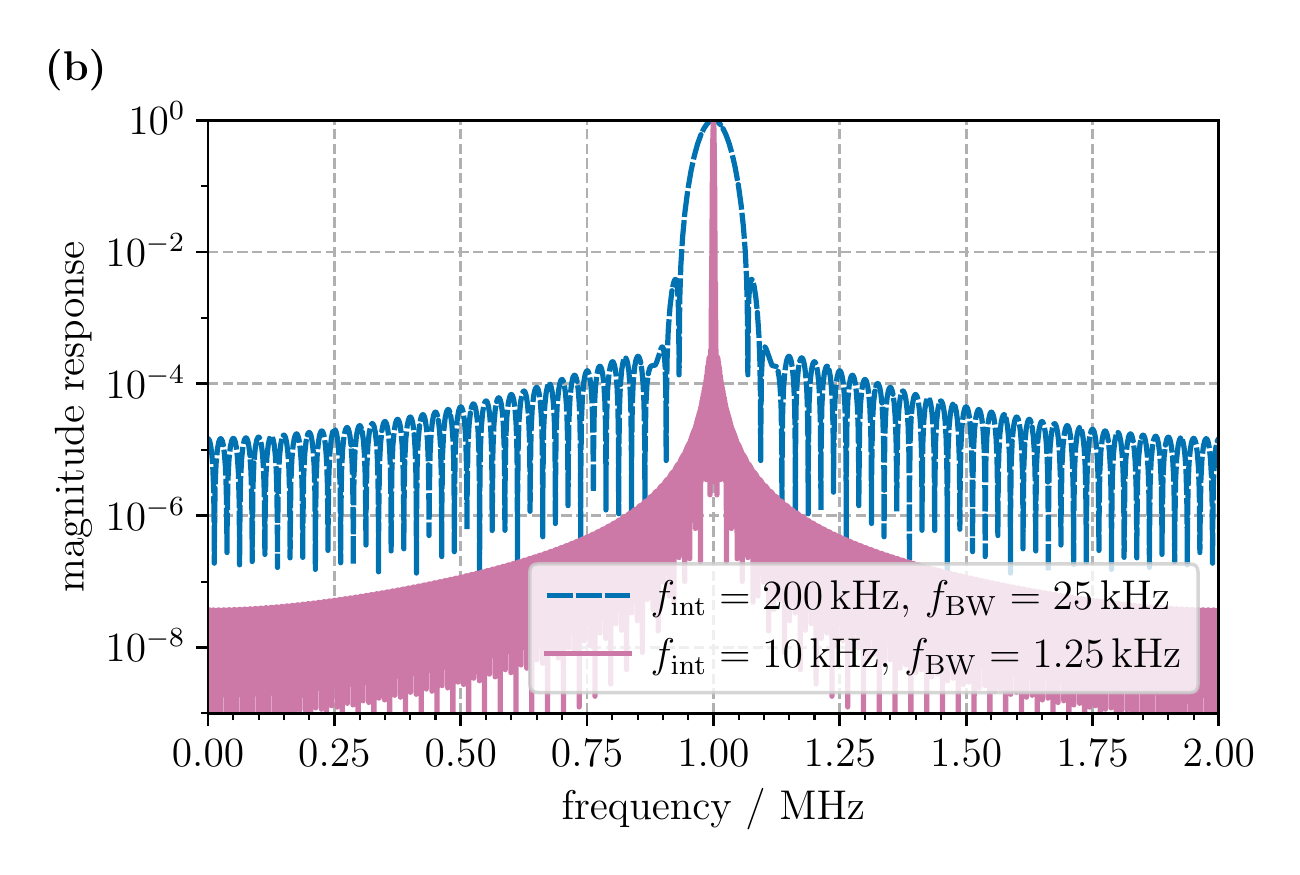}
	\caption{(a) Example of coefficients for \glsentryshort{iq} demodulation and filtering. (b) Frequency response of the demodulation filter for two intermediate frequency values. }
	\label{fig:filter-response}
\end{figure}

The demodulation \gls{fir} filter is obtained combining a low-pass Hamming window with the \gls{iq} references, skipping the samples corresponding to \gls{fir} coefficients equal to zero. An example is shown in Fig. \ref{fig:filter-response}(a). The filter order is limited by the \gls{mcu} speed, with $8 f_\text{smp}/f_\text{int}$ filter coefficients. The filter bandwidth is designed as $f_\text{BW}=f_\text{int}/8$. Different configurations, corresponding to $f_\text{int}$ between \SI{10}{\kilo\hertz} and \SI{200}{\kilo\hertz}, are pre-loaded in the \gls{mcu} memory, to allow on-the-fly reconfiguration of the demodulation stage via software. Examples of filter magnitude responses are reported in Fig. \ref{fig:filter-response}(b). The optimal configuration in terms of dataflow processing stability and phase unwrap efficiency is typically obtained for $f_\text{int}=\SI{100}{\kilo\hertz}$.

The coefficients of the \gls{fir} \gls{lpf} for the output data decimation to rate $f_\text{out}$ are calculated considering an Hamming window, with $12 f_\text{int}/f_\text{out}$ coefficients, and bandwidth close to $f_\text{out}/3$, optimized to reduce aliasing due to side-lobes of the filter transfer function. The coefficients are pre-calculated for $f_\text{out}$ between \SI{500}{\hertz} and \SI{20}{\kilo\hertz}.

The \gls{lpf} for the frequency drift correction is a first-order \gls{iir} kind. The filter and \gls{pid} parameters can be configured by software, depending on the drift dynamics. The cutoff frequency is typically kept very low in order not to interfere with the signal high-frequency components of interest.

\subsubsection{Synchronization}
\label{sec:synch}

In the case of synchronization of remote systems, with the trigger event being encoded on the signal to be measured, the choice of $f_\text{int}$ impacts on the achievable synchronization resolution. Therefore during the triggering procedure $f_\text{int}$ is set at the highest possible value, i.e. $f_\text{int}=\SI{200}{\kilo\hertz}$. In this configuration, the maximum jitter between independent trigger instances is $\pm\SI{5}{\micro\second}$. Assuming an uniform uncertainty distribution, the expected standard deviation over the initial timescale is $\sigma_\text{sync} \sim \SI{3}{\micro\second}$.

After the trigger event is detected, each board switches to the desired $f_\text{int}$ value and starts the acquisition. The loss of synchronization will depend solely on the frequency offset and drift for the independent clock sources, mostly due to temperature fluctuations and quartz aging. A firmware routine is dedicated to resynchronize the acquisitions on-the-fly, with a small latency of about \SI{110}{\micro\second}. The routine combines the trigger detection with a measurement of the relative delay accumulated since the last resynchronization event. This allows to compensate for the oscillator frequency fluctuations, fine-tuning the \gls{ocxo} adjust voltage on a daily timescale.

The residual delay accumulated by the independent \glspl{ocxo} can be estimated assuming a linear fractional frequency drift, which has been measured to be $D \sim \SI{2e-13}{\per\second}$. Accordingly, the time interval after which the cumulative clock delay reaches the time accuracy of $\sigma_\text{sync}$ is $\sqrt{2\sigma_\text{sync}/D} \sim \SI{5000}{\second}$. This means that performing a resynchronization procedure every hour is sufficient to maintain the timescales synchronized to within $\sigma_\text{sync}$.

\subsubsection{Numerical resolution}

Before unwrapping, $\delta \phi'_m[n]$ is defined between $-2\pi$ and $+2\pi$. This means that the numerical resolution for calculations with 32-bit floating-point variables is always better than \SI{0.5}{\micro\radian}. This corresponds to a frequency resolution of \SI{8}{\milli\hertz} at the typical intermediate frequency $f_\text{int}=\SI{100}{\kilo\hertz}$, stressing that this is a conservative estimation related to big $\Delta\nu$ values. Amplitude is defined between \SI{0}{\volt} and $V_\text{ref}/2 = \SI{1.25}{\volt}$, hence its 32-bit floating-point numerical resolution is better than \SI{0.1}{\micro\volt}.

For transmission bandwidth optimization, the phase and amplitude samples are transmitted to the control unit as integers, with 32-bit and 16-bit resolution respectively, corresponding to \SI{1.5}{\nano\radian} (or \SI{23}{\micro\hertz} with $f_\text{int}=\SI{100}{\kilo\hertz}$) and \SI{19}{\micro\volt}. Effects of numerical truncation can be expected beyond these limits.

\subsection{Software control}

The acquired data are transmitted in real-time to a computer via \gls{usb} connection. The decimated data are packed in chunks containing incremental phase, amplitude, frequency drift correction, and other monitoring variables, with an output data rate up to \SI{20}{\kilo\hertz}. The data are received by a software written in Python, checked for \gls{crc} integrity and stored incrementally in HDF5 binary files, together with the system settings. The software controls the embedded system functions via serial commands, and allows to configure its parameters on-the-fly.

\section{Experimental characterization}

\begin{figure*}[!t]
	\centering
	\includegraphics[width=1\columnwidth]{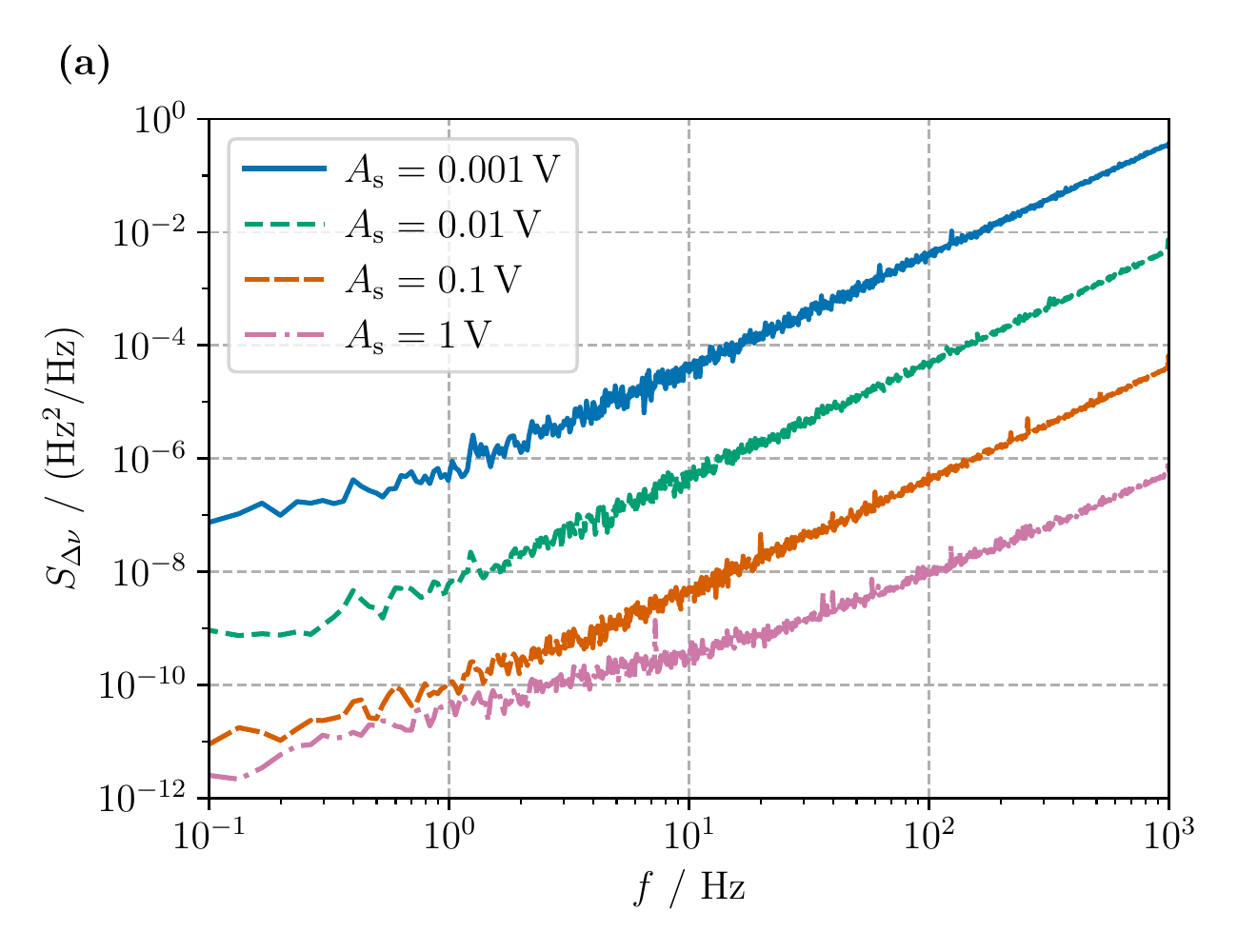}
	\includegraphics[width=1\columnwidth]{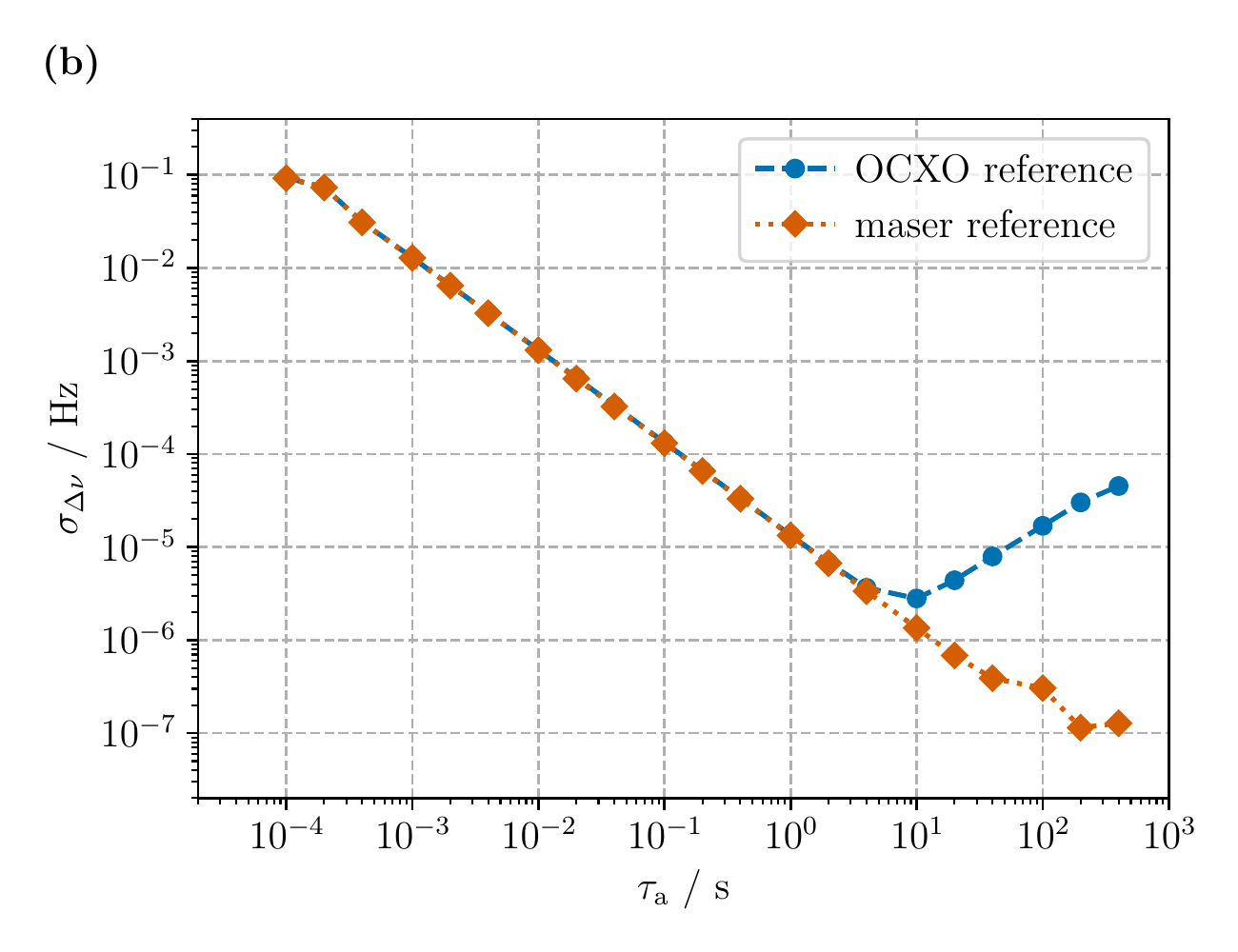}\\
	\includegraphics[width=1\columnwidth]{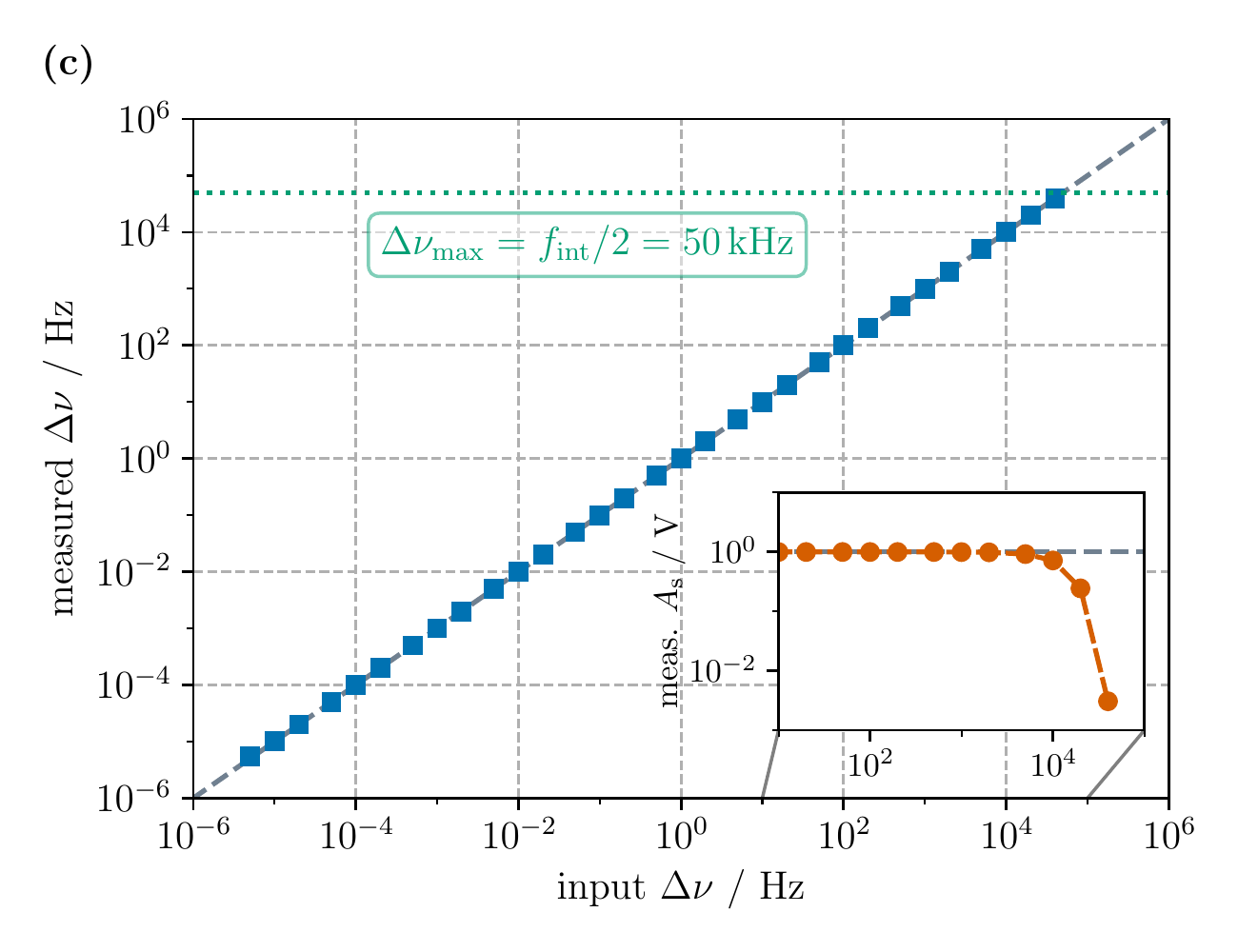}
	\includegraphics[width=1\columnwidth]{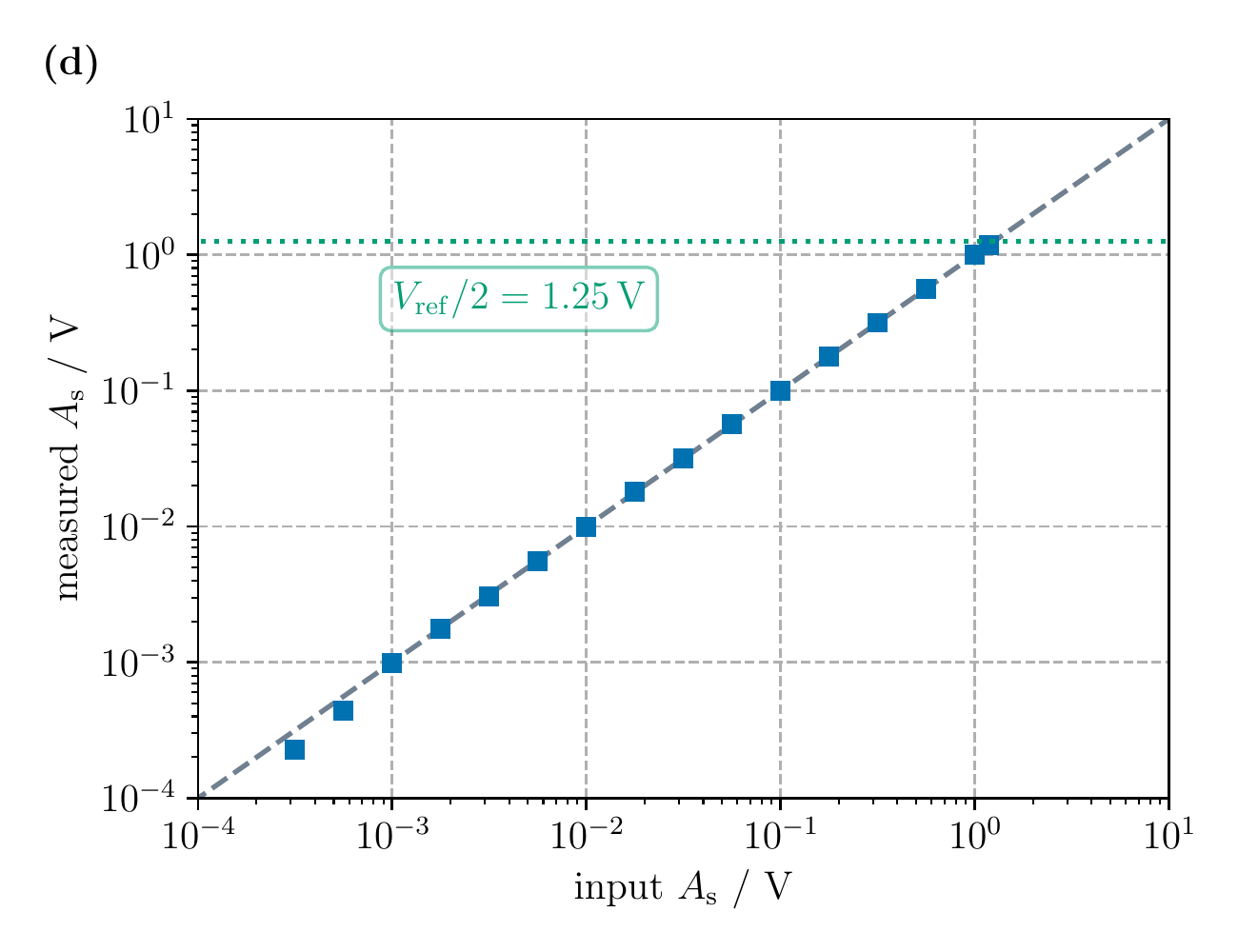}
	\caption{(a) \glsentryshort{psd} of frequency signals acquired with the measurement board as a function of Fourier frequency; the signals are generated with \SI{1}{\mega\hertz} carrier frequency and variable amplitude. (b) Allan deviation calculated for \SI{1}{\mega\hertz} signals at \SI{1}{\volt} amplitude as a function of averaging time, considering either maser or \glsentryshort{ocxo} reference for the board clock. (c) Measured frequency as a function of nominal input frequency at \SI{1}{\volt} amplitude; the dashed diagonal represents the expected correspondence; the inset shows the corresponding measured amplitude. (d) Amplitude measured as a function of nominal input amplitude for \SI{1}{\mega\hertz} signals; the dashed diagonal represents the expected correspondence.}
	\label{fig:noise-linearity}
\end{figure*}

Our measurement system has been characterized with \gls{rf} signals generated by a commercial synthesizer (3352A, Agilent), externally referenced to a hydrogen maser. The board was also referenced to the same stable source, to exclude noise contributions due to the \gls{ocxo}. The measurement noise was estimated by acquiring sinusoidal signals having \SI{1}{\mega\hertz} nominal frequency. The acquisition was configured with $f_\text{int}=\SI{100}{\kilo\hertz}$ and $f_\text{out}=\SI{4}{\kilo\hertz}$. Fig. \ref{fig:noise-linearity}(a) shows the \gls{psd} $S_{\Delta\nu}(f)$, with $f$ the Fourier frequency, of the frequency fluctuations $\Delta\nu$ measured at different signal amplitudes, up to $A_\text{s}=\SI{1}{\volt}$. The noise spectra is characterized by white phase noise, with a negligible additional noise in the low-frequency region. Weaker signals exhibit higher phase noise floors, due to the lower signal-to-noise ratio in the demodulation band: notably the noise power scales as $1/A_\text{s}^2$. This is consistent with the \gls{adc} noise expected considering the resolution given by its \gls{enob}.

Fig. \ref{fig:noise-linearity}(b) shows the frequency instability calculated in terms of the overlapping Allan deviation $\sigma_{\Delta\nu}(\tau_\text{a})$, with $\tau_\text{a}$ the averaging time, for a \SI{1}{\mega\hertz} signal at amplitude \SI{1}{V}, acquired for \SI{2000}{\second} with $f_\text{int}=\SI{100}{\kilo\hertz}$ and $f_\text{out}=\SI{10}{\kilo\hertz}$. The curve exhibits the expected behavior for white-phase-limited measurements on a bandwidth of about $\SI{4}{\kilo\hertz}$. In particular, the impact of white phase noise on the Allan deviation could be reduced by reconfiguring the output filter to a lower measurement bandwidth. This considerably mitigates the aliasing effects which are often encountered with frequency counters \cite{calossoAvoiding2016}. When the measurement is repeated with the acquisition system referenced to the onboard \gls{ocxo}, the quartz drift becomes dominant for averaging times above~$\sim \SI{10}{\second}$. 

When $f_\text{int}=\SI{100}{\kilo\hertz}$, the maximum frequency that can be measured without aliasing is $f_\text{int}/2=\SI{50}{\kilo\hertz}$; actually, the demodulation filter has a cutoff of $f_\text{BW}=\SI{12.5}{\kilo\hertz}$, and acts as anti-aliasing filter. To address the frequency response of the demodulation, signals at different carrier frequencies $\nu_0+\Delta \nu$ were acquired, with $\Delta \nu$ between \SI{5}{\micro\hertz} and \SI{40}{\kilo\hertz}, and amplitude fixed to \SI{1}{\volt}. The measured frequency was averaged over \SI{50}{\second} at $f_\text{out}=\SI{10}{\kilo\hertz}$, and compared to the nominal input frequency as shown in Fig. \ref{fig:noise-linearity}(c). A linear fit to the data allows quantifying the measurement linearity, as well as an assessment of the demodulation process accuracy over \num{10} orders of magnitude. The linearity coefficient is \num{1} within \num{2e-8}, while the intercept of about \SI{20}{\micro\hertz} can be attributed to numerical resolution and truncation. The inset of Fig. \ref{fig:noise-linearity}(c) quantifies the signal amplitude attenuation as a function of the input frequency: it confirms that frequency components above \SI{30}{\kilo\hertz} are strongly suppressed by the demodulation filter, by more than \SI{20}{\decibel}. Aliasing effects at the phase unwrap stage are therefore negligible.

We characterized the linearity in the amplitude demodulation by acquiring signals with fixed $\Delta \nu=\SI{0}{\hertz}$ and variable amplitude, between \SI{0.3}{\milli\volt} and \SI{1.2}{\volt}. The maximum amplitude that can be measured without distortion is half of the \gls{adc} voltage reference, hence $V_\text{ref}/2=\SI{1.25}{\volt}$. The results are shown in Fig. \ref{fig:noise-linearity}(d). The corresponding linearity coefficient depends on the knowledge of the input amplification gain, which is about \num{15}. Nonetheless, the linearity relative uncertainty of \num{7e-4} and the bias of \SI{0.3}{\milli\volt} give figures for the corresponding amplitude resolution over almost \num{4} orders of magnitude. We also verified that cross-talks between phase and amplitude are negligible, as long as the signal carrier lays within the demodulation filter band.

\begin{figure}[!t]
	\centering
	\includegraphics[width=1\columnwidth]{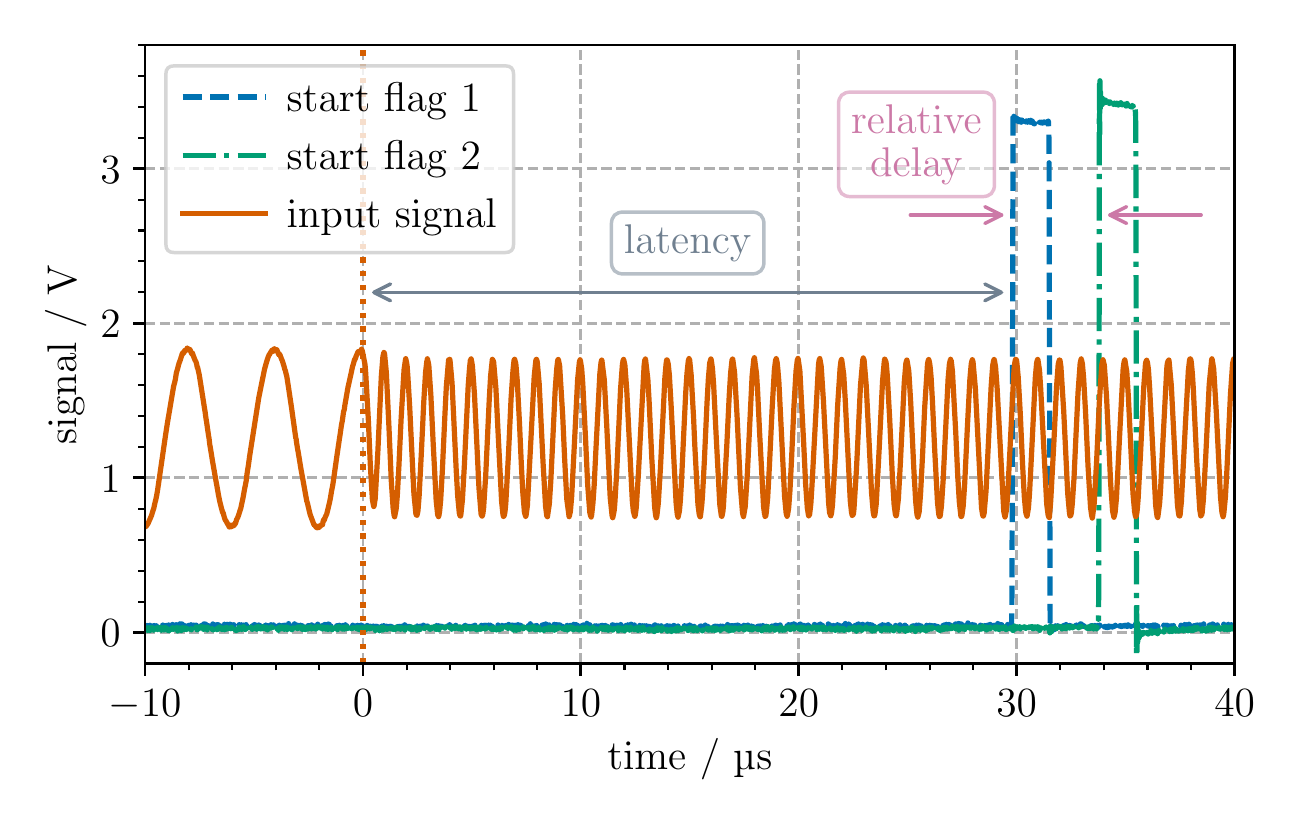}
	\caption{Signals related to the acquisition with two synchronized boards, acquired with an oscilloscope. A common input \glsentryshort{rf} signal is split and sent to each system. The carrier frequency is shifted to \SI{1}{\mega\hertz} at time \SI{0}{\second}, and each board, $1$ and $2$, starts the acquisition after a latency interval.}
	\label{fig:signals-sync}
\end{figure}

The mechanism of remote synchronization was tested by arming two different boards to wait for a trigger signal with amplitude greater than half the nominal amplitude. A common input signal was obtained through down-conversion using the \gls{dds}, split and sent to the two boards. The signal frequency was stepped between \SI{1.25}{MHz} and \SI{1}{MHz}, to bring it within the demodulation band and generate a rising amplitude edge, acting as trigger. The acquired signal was monitored with an oscilloscope, as shown in Fig. \ref{fig:signals-sync}. The triggering instants on each board were detected by raising digital output flags when the acquisitions started. An average latency of \SI{32}{\micro\second} was measured between the instant at which the carrier is shifted into the demodulation band and the actual acquisition start. This latency depends only on the system architecture, since it is caused by the deterministic initialization time required for the \gls{mcu} peripherals and by the \gls{dsp} algorithm, e.g., the intrinsic \gls{fir} filter delay. The average relative delay between the two boards over \num{30} synchronized acquisitions showed a synchronization uncertainty given by a standard deviation of $\sigma_\text{sync} = \SI{2.9}{\micro\second}$, as expected for $1/f_\text{int}=\SI{5}{\micro\second}$ and discussed in Section \ref{sec:synch}.

\section{Optical phase sensing}

We now consider the case study of two ultrastable lasers placed in two distant locations, connected by a telecom fiber link established on a pair of unidirectional optical fibers, running parallel inside the same cable. At each location we interfere the laser signals, we measure the beat-note frequency, and we synchronize the respective acquisition platforms by encoding a trigger event on the ultrastable carriers. We compensate in real-time the slow frequency drifts of the beat-note, and we demonstrate the possibility of measuring both the fiber and laser noises by suitably combining the synchronized acquisitions.

\subsection{Optical setup}

\begin{figure}[!t]
	\centering
	\includegraphics[width=1\columnwidth]{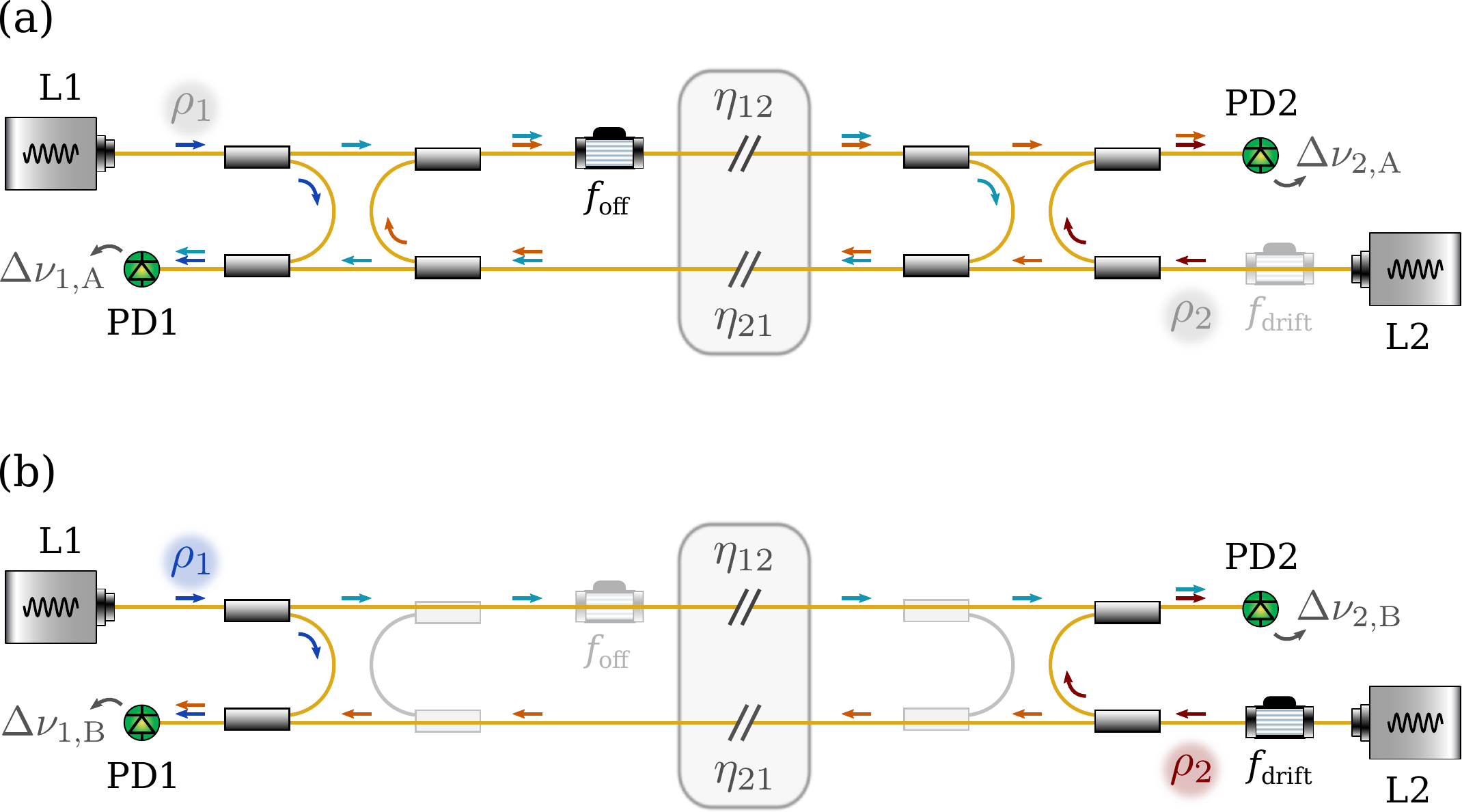}
	\caption{Optical schemes for the comparison between two ultrastable lasers L1 and L2 through a fiber link. Fiber splitters are used to split and combine signals; \glsentryshortpl{aom} are used to shift the carrier frequency by $f_\text{off}$ and $f_\text{drift}$. The laser signals interfere on photodiodes PD1 and PD2, and the beat-notes are acquired with independent systems to extract the relative laser noise $\rho_2 - \rho_1$ and the fiber noise $\eta$. (a) Self-heterodyne detection. (b) Heterodyne detection.}
	\label{fig:scheme-optical}
\end{figure}

We implemented an experimental layout that allows either the self-heterodyne interference of each laser source independently, exploiting the round-trip path in the fiber as a delay line, or the heterodyne interference of the two lasers on a two-way approach. The optical setup is shown in Fig. \ref{fig:scheme-optical}.

Each laser source (L1 and L2) is a commercial external cavity diode laser with \SI{1542}{nm} wavelength and intrinsic \SI{10}{kHz} linewidth, frequency stabilized to a \SI{5}{\centi\meter} long, high-finesse cavity, using the Pound-Drever-Hall technique \cite{blackIntroduction2001}, on a bandwidth of \SI{200}{kHz}. The resulting short-term stability is \num{2e-15}, with a drift of the order of \SI{0.5}{\hertz\per\second}.

The diode lasers are injected into single-mode optical fibers. Starting from L1, \SI{5}{\percent} of the laser light is split and used as local reference. The remainder \SI{95}{\percent} is sent to the remote location through the fiber. The link is \SI{36}{km} long and covers both city and country areas, with a one-way light travel time of $\tau \simeq \SI{180}{\micro\second}$. At the remote location, the incoming light from L1 is split using a 50:50 coupler: half of the power is interfered with the L2 laser reference using a 50:50 coupler, the rest is routed back to the transmitting source, where it interferes with the L1 reference light. The light from L2 follows a symmetrical path.

An \gls{aom} driven at $f_\text{off} = \SI{40}{\mega\hertz}$ is introduced on the fiber connecting L1 and L2, close to L1. Another \gls{aom}, driven at adjustable frequency $f_\text{drift}$, is placed at the L2 output, and it will be used as actuator for the carrier frequency drift correction.

The beat-notes are acquired on each side with high-speed photodiodes, PD1 and PD2. Two interference signals corresponding to two optical layouts are simultaneously present on each photodiode: self-heterodyne detection from the self-delayed interference of each laser, and heterodyne detection from the interference between L1 and L2. In the self-heterodyne scheme the carrier frequency is determined by $f_\text{off}= \SI{40}{\mega\hertz}$. In the heterodyne configuration each photodiode detects the beat-note between the reference signal and the incoming light from the remote laser: in our experiment, this is about \SI{1.1}{GHz}. Accordingly, thanks to the spectral separation of the self-heterodyne and heterodyne beat-notes, the measurement scheme can be easily reconfigured to process one or the other by changing the \gls{lo} frequency $f_\text{LO}$ at the down-conversion stages, i.e., to shift the \gls{rf} signal of interest to the nominal frequency $\nu_0=\SI{1}{\mega\hertz}$. The down-converted signals are acquired and demodulated by the respective measurement boards.

The optical phase is mainly affected by two kinds of noise contributions: the intrinsic noise of the laser sources, $\rho_1$ and $\rho_2$ respectively, and the noise accumulated along the optical fiber link $\eta$, typically introduced by seismic or anthropic vibrations, acoustic noise, physical deformations, strains, and temperature variations. The measurement of these noise signals for the two considered detection schemes will be discussed in the following paragraphs.

\subsection{Self-heterodyne detection}

Let us consider the self-heterodyne scheme of Fig. \ref{fig:scheme-optical}(a), where each laser interferes with its self-delayed light. The beat-notes at frequency $f_\text{off}$ are down-converted to $\nu_0$ using synthesizers set to $f_\text{LO}=\SI{41}{MHz}$. In this configuration the drift correction \gls{aom} is kept at fixed frequency, hence $f_\text{drift}$ is neglected. The frequency fluctuations acquired by photodiodes PD1 and PD2 at the respective locations of L1 and L2 are
\begin{subequations}
	\label{eq:self-noise-full}
	\begin{align}
	\begin{split}
		\Delta \nu_\text{1,A}(t_1) & =  \rho_1(t_1-2\tau) -\rho_1(t_1) \\
		& +\eta_{12}|_{t_1-2\tau}^{t_1-\tau} + \eta_{21}|_{t_1-\tau}^{t_1} + \Delta f_\text{off}(t_1-2\tau) 
	\end{split}\\
	\begin{split}
		\Delta \nu_\text{2,A}(t_2) & =  \rho_2(t_2-2\tau) - \rho_2(t_2) \\
		& +\eta_{21}|_{t_2-2\tau}^{t_2-\tau} + \eta_{12}|_{t_2-\tau}^{t_2} + \Delta f_\text{off}(t_2-\tau)\,. 
	\end{split}
	\end{align}
\end{subequations}
Here $t_1$ and $t_2$ represent the local timescales, which are natively independent, $\rho_1(t_1)$ is the L1 laser frequency noise at time $t_1$, $\eta_{21}|_{t_1-\tau}^{t_1}$ is the one-way fiber frequency noise integrated over the path from L2 to L1 in the interval between $t_1-\tau$ and $t_1$, and correspondingly for the other terms. $\Delta f_\text{off}$ is the \gls{aom} frequency deviation from its nominal value \SI{40}{\mega\hertz}.

The acquisition timescales $t_1$ and $t_2$ are synchronized exploiting the triggering mechanism described in Section \ref{sec:synch}. A step frequency shift is imprinted on $f_\text{off}$, driven by the onboard synthesizer: the self-heterodyne beat-note at each terminal is momentarily shifted out of the detection band, and the resulting rising amplitude edge is used to trigger the two acquisitions. After the synchronization process, a fixed time-relation can be established between the remote boards. This is defined as $t=t_1-\tau=t_2$, where the relative delay $\tau$ accounts for the fact that the synchronizing pulse travels along the fiber before being detected at the other end. Variations of the propagation path delay are sufficiently smaller than the intrinsic synchronization uncertainty $\sigma_\text{sync}$, hence they can be ignored for the applications considered here. Therefore, after synchronization, the acquired signals are
\begin{subequations}
	\label{eq:self-noise-sync}
	\begin{align}
		\Delta \nu_\text{1,A}(t) = & \rho_1(t-2\tau) -\rho_1(t) + \eta_{12}|_{t-2\tau}^{t-\tau} + \eta_{21}|_{t-\tau}^{t}\\
		\Delta \nu_\text{2,A}(t) = & \rho_2(t-2\tau) - \rho_2(t) + \eta_{12}|_{t-\tau}^{t} + \eta_{21}|_{t-2\tau}^{t-\tau}\,.
	\end{align}
\end{subequations}
The \gls{aom} frequency fluctuations after the synchronization routine are considered negligible with respect to the other terms, and neglected through the rest of the analysis.

It can be assumed that, to a first approximation, the laser noise cancels out, hence $\rho(t) \approx \rho(t-2\tau)$, which is valid for frequency fluctuations with spectral components $f \ll 1/\tau$. With $\tau \simeq \SI{180}{\micro\second}$, this is justified for spectral analysis below \SI{1}{\kilo\hertz} Fourier frequency. Moreover, the phase perturbations $\eta_{12}$ and $\eta_{21}$ are expected to be mostly correlated, as the noise of the two fibers, laying in the same cable, is common to the first order. It follows that the interference signals are mostly affected by the common fiber noise $\eta \approx \eta_{12} \approx \eta_{21}$, and the round-trip frequency fluctuations can be approximated as
\begin{subequations}
	\label{eq:self-noise-approx}
	\begin{align}
		\Delta\nu_\text{1,A}(t) \simeq & 2 \eta(t)\\
		\Delta\nu_\text{2,A}(t) \simeq & 2 \eta(t)\,.
	\end{align}
\end{subequations}

\begin{figure}[!t]
\centering
\includegraphics[width=1\columnwidth]{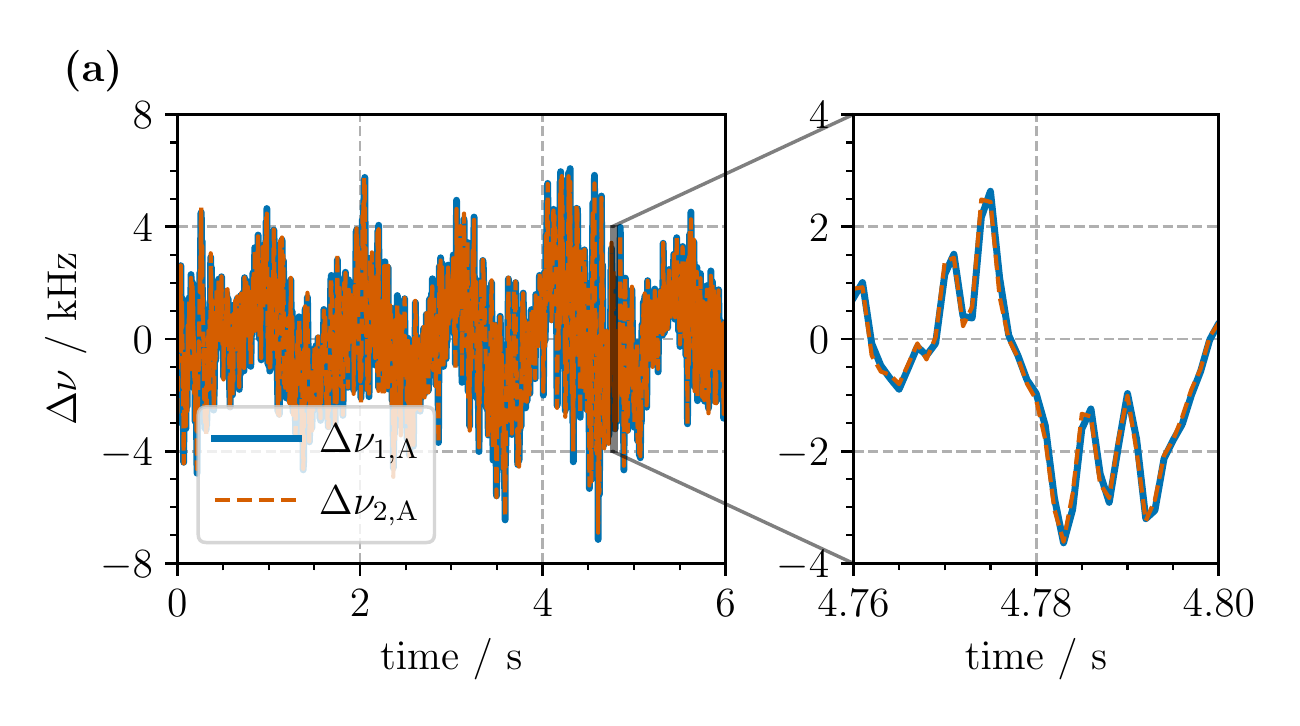}
\includegraphics[width=1\columnwidth]{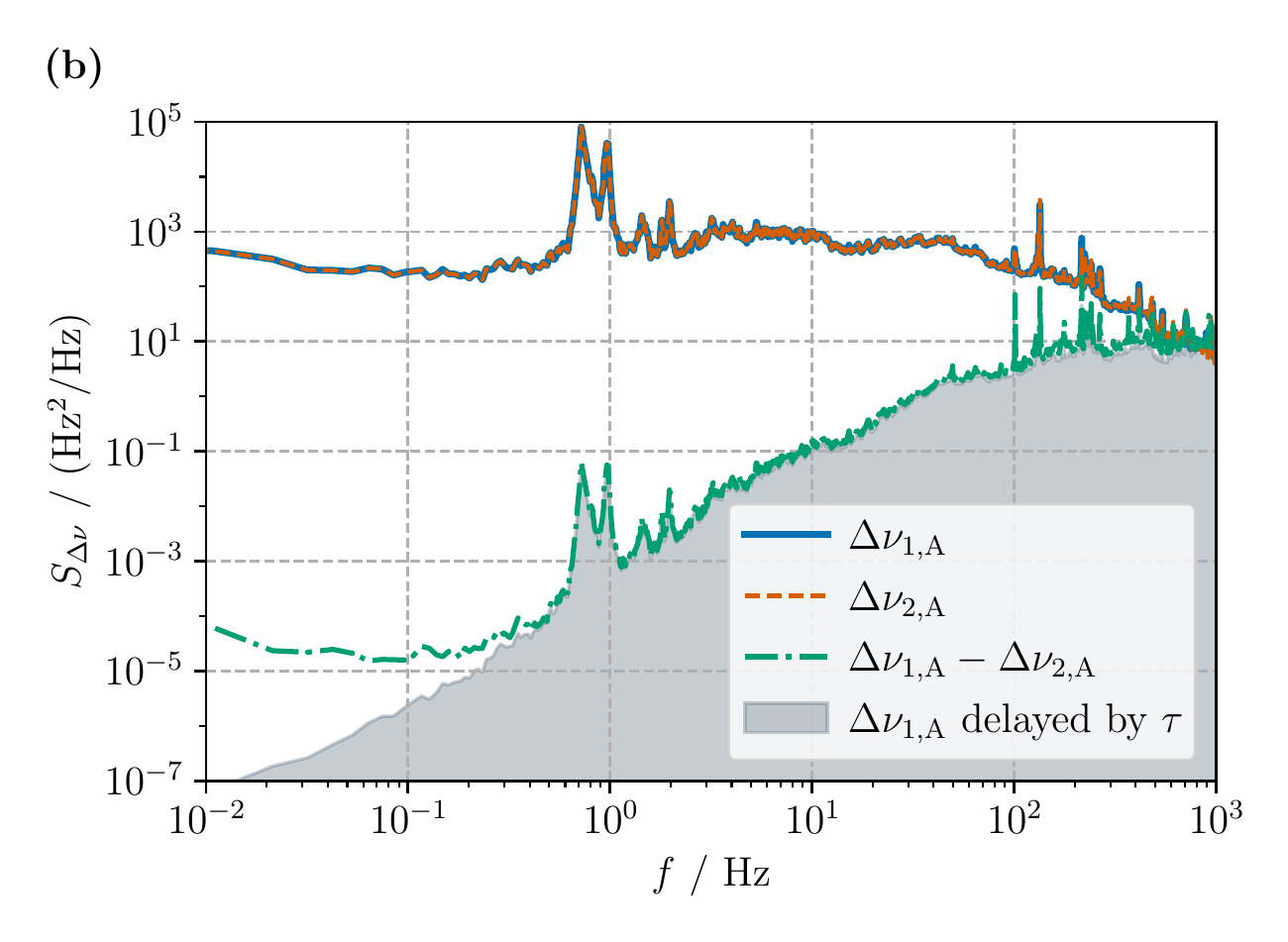}
\caption{(a) Synchronized acquisitions of frequency fluctuations over the optical fiber link, in the self-heterodyne interference scheme; the tracks mostly overlap. (b) \glsentryshort{psd} of the acquired signals; the noise spectrum of their difference is also shown, compared with the noise expected for a synchronization delay $\tau \simeq \SI{180}{\micro\second}$.}
\label{fig:scpectra-self}
\end{figure}

The degree of temporal correlation, as well as the quality of the synchronization process, can be appreciated from Fig. \ref{fig:scpectra-self}(a), showing the coincidence between frequency fluctuation signals acquired at \SI{1}{\kilo\hertz} by the two distant boards. Fig. \ref{fig:scpectra-self}(b) shows the \gls{psd} of the frequency noise for each of the two beat-notes, calculated for \SI{50}{\minute} long acquisitions. The spectra mostly overlap, since from \eqref{eq:self-noise-approx} they refer to the same physical signals. The measurements are consistent with the noise expected for an urban fiber link, where the phase perturbations generated by different anthropic, seismic, oscillatory, or acoustic sources accumulate along the fiber path. Indeed, the measured white frequency noise level is compatible with other land-deployed cables reported in literature \cite{akatsuka30kmlong2014,akatsukaOptical2020,lopezCascaded2010}, assuming a linear scaling of the noise \gls{psd} with the link length. Nevertheless, the actual scaling strongly depends on the specific deployment, which can cross data centers, bridges, buildings and suspended fiber segments. The observation of daily fluctuations in the noise spectrum, \SIrange{5}{15}{\decibel} higher during the day compared to night, can be attributed to human activities, such as vehicle traffic along the roads adjacent to the fiber path. These aspects play an important role in applications where the dissemination of stable phase signals is required \cite{williamsHighstability2008}.

The spectrum of the difference between the two synchronized acquisitions is also shown in Fig. \ref{fig:scpectra-self}(b). The residual noise level that is observed for $\Delta\nu_\text{1,A}-\Delta\nu_\text{2,A}$ is consistent with the expectations for the difference between correlated, yet delayed, signals. Indeed, from \eqref{eq:self-noise-sync}, the acquisitions have a relative delay equal to the light travel time $\tau$, given by the synchronization procedure, and \cite{calossoFrequency2014}, the spectrum of the round-trip fiber noise difference can be estimated as $(2\pi\tau f)^2 S_{\Delta\nu_\text{1,A}}(f)$, with $S_{\Delta\nu_\text{1,A}}(f)$ the \gls{psd} of $\Delta\nu_\text{1,A} \simeq \Delta\nu_\text{1,B}$ at Fourier frequency $f$. The measurement deviates from the expectations only at very low frequencies, where other effects emerge. We attribute this additional noise to optical length variations for the short fibers which are not common for the two path.

\subsection{Heterodyne detection}

In the heterodyne configuration, shown in Fig. \ref{fig:scheme-optical}(b), each local laser interferes with the light coming from the remote laser. The beat-notes between L1 and L2 are down-converted to $\nu_0$ by setting $f_\text{LO}$ to about \SI{1.1}{\giga\hertz}. The \gls{aom} at $f_\text{off}$ is driven at fixed frequency, hence it is neglected in this scheme. The optical carrier of L2 is shifted exploiting the \gls{aom} driven at $f_\text{drift}$ by the onboard \gls{dds}, centered at \SI{40}{\mega\hertz}, to correct the slow beat-note drifts that are due to uncorrelated length variations of the two optical cavities. Similarly to the self-heterodyne case, in the heterodyne scheme the acquisitions are synchronized by imprinting a step frequency shift on $f_\text{drift}$, defining the common timescale $t=t_1-\tau=t_2$. The following frequency fluctuations are expected to affect the PD1 and PD2 photodiode signals:
\begin{subequations}
	\begin{align}
	\begin{split}
		\Delta\nu_\text{1,B}(t) = & \rho_2(t-\tau) - \rho_1(t) \\&+ \Delta f_\text{drift}(t-\tau) + \eta_{21}|_{t-\tau}^{t}
	\end{split}\\
	\begin{split}
		\Delta\nu_\text{2,B}(t) = & \rho_1(t-\tau) - \rho_2(t) \\&- \Delta f_\text{drift}(t) + \eta_{12}|_{t-\tau}^{t} \,.
	\end{split}
	\end{align}
\end{subequations}

The drift correction frequency results from the \gls{pid} implemented in the \gls{mcu} firmware, as described in Section \ref{sec:drift}. This tracks the low frequency components of $\Delta\nu_\text{2,B}$, with a \gls{lpf} cutoff of \SI{0.01}{\hertz}. In that spectral region the relative drift between the independent lasers L1 and L2 dominates, hence $\Delta f_\text{drift}$ mainly follows the low-frequency components of $\rho_1 - \rho_2$. Due to symmetry, any frequency offset cancels out for both $\Delta\nu_\text{1,B}$ and $\Delta\nu_\text{2,B}$.

Although in metrology the fiber link noise represents a limit for the comparison of optical frequency references and requires phase stabilization techniques \cite{williamsHighstability2008}, from a different perspective the fiber can be used as a sensor for earthquakes and other geophysical events occurring close to the fiber \cite{marraOptical2022}. With the heterodyne scheme, the reference laser and fiber noise contributions can be distinguished. Under the same assumptions introduced for \eqref{eq:self-noise-approx}, the approximated frequency fluctuations on the two photodiodes become
\begin{subequations}
\label{eq:fiber-hetero-approx}
\begin{align}
	\Delta\nu_\text{1,B}(t) \simeq & \rho_2(t) - \rho_1(t) + \Delta f_\text{drift}(t) + \eta(t) \\
	\Delta\nu_\text{2,B}(t) \simeq & \rho_1(t) - \rho_2(t) - \Delta f_\text{drift}(t) + \eta(t)\,.
\end{align}
\end{subequations}
It follows that, from the combination of $\Delta\nu_\text{1,B}$ and $\Delta\nu_\text{2,B}$, it is possible to extract the fiber noise
\begin{equation}
	\label{eq:fiber-noise-hetero}
	\Delta\nu_\text{1,B}(t)+\Delta\nu_\text{2,B}(t) \simeq 2\eta(t)
\end{equation}
and the relative laser noise
\begin{equation}
	\label{eq:laser-noise-hetero}
	\Delta\nu_\text{1,B}(t)-\Delta\nu_\text{2,B}(t) - 2\Delta f_\text{drift}(t) \simeq 2(\rho_2(t) - \rho_1(t))\,.
\end{equation}

\begin{figure}[!t]
\centering
\includegraphics[width=1\columnwidth]{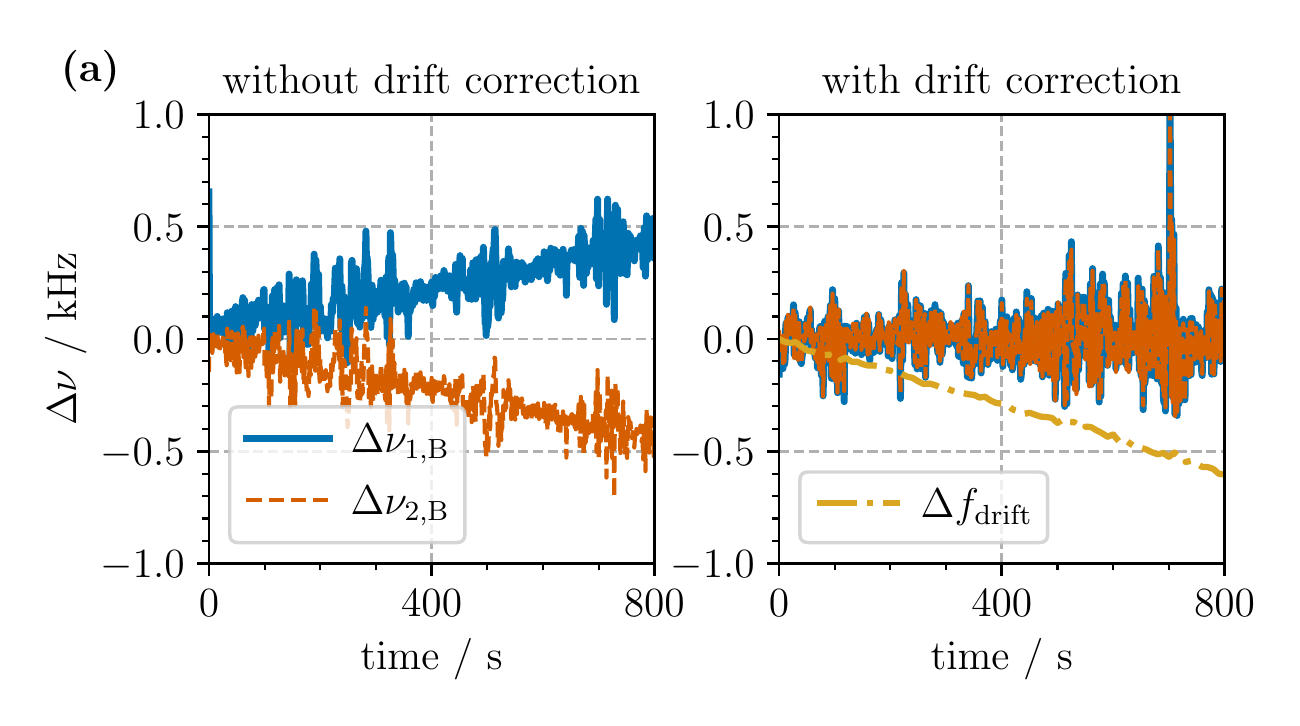}
\includegraphics[width=1\columnwidth]{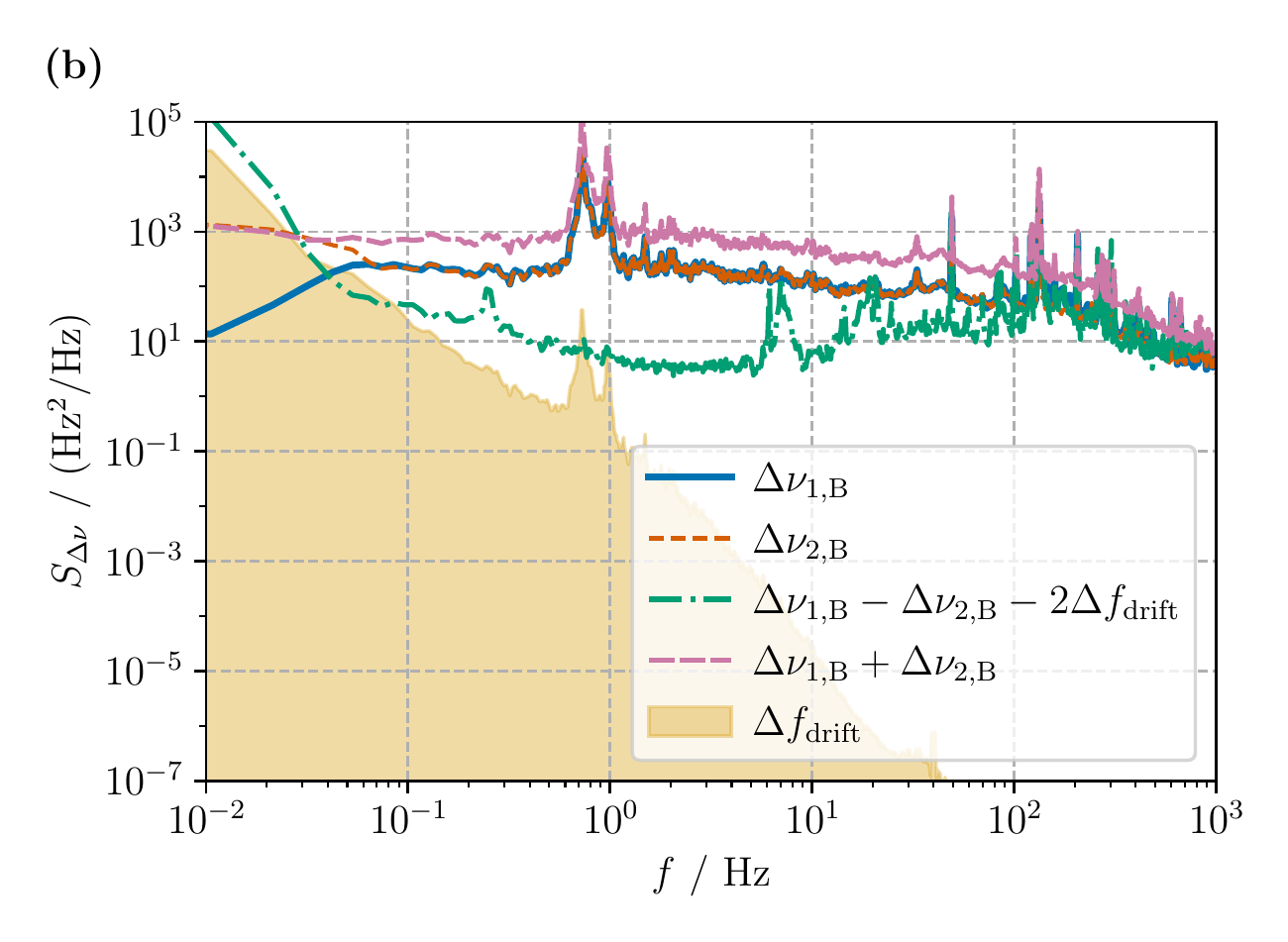}
\caption{(a) Synchronized acquisitions of frequency fluctuations over the optical fiber link, in the heterodyne interference scheme, without and with frequency drift correction; with drift correction the tracks mostly overlap. (b) \glsentryshort{psd} of the acquired signals, with drift correction; the noise spectra of their sum and difference are also reported.}.
\label{fig:scpectra-hetero}
\end{figure}

The effects of frequency drifts can be seen from Fig. \ref{fig:scpectra-hetero}(a). Without activating the drift correction, $\Delta\nu_\text{1,B}$ and $\Delta\nu_\text{2,B}$ show significant offsets, variable in time. When the drift correction is enabled, the synchronized tracks are characterized by fast frequency fluctuations that lay around zero, with the low-frequency drift being tracked by $\Delta f_\text{drift}$.

The \glspl{psd} of the acquired signals are shown in Fig. \ref{fig:scpectra-hetero}(b), mostly overlapping except at low frequencies, where $\Delta f_\text{drift}$ dominates. $\Delta\nu_\text{1,B}$ and $\Delta\nu_\text{2,B}$ are combined to show either the fiber noise or the relative laser frequency noise using \eqref{eq:fiber-noise-hetero} and \eqref{eq:laser-noise-hetero}. It can be seen that the fiber noise represents the main contribution in the measurements for our setup, and that it is similar to the self-heterodyne results presented in Fig. \ref{fig:scpectra-self}(b). The measured laser noise is consistent with the expectations for ultrastable sources \cite{clivatiPlanarwaveguide2011}, noting that the lasers are installed in suboptimal conditions, outside a laboratory and subject to environmental disturbances. For example, some acoustic spurs are present above \SI{10}{\hertz}, likely given by cooling fans close to the setups. The laser noise dominates at very low frequencies below \SI{0.1}{\hertz}, where the laser drift is compensated by the frequency drift correction.

\section{Conclusion}

In the current work, a phase-sensitive detector has been designed and implemented on a \gls{mcu}. Dedicated electronic and firmware platforms have been realized and characterized. Following the approach of digital lock-in amplifiers, the system implements efficient algorithms for phase and amplitude heterodyne demodulation of \gls{rf} signals down-converted to \SI{1}{\mega\hertz}. An output data rate of \SI{20}{\kilo\hertz} can be achieved, with a reconfigurable demodulation bandwidth between \SI{1.25}{\kilo\hertz} and \SI{12.5}{\kilo\hertz}. The input frequency deviation from the \SI{1}{\mega\hertz} carrier can be measured over almost \num{10} orders of magnitude, from a few \si{\micro\hertz} up to \SI{50}{\kilo\hertz}. The accepted input signal amplitude spans over more than \num{3} voltage decades. Frequency instability at \SI{1}{\second} is about \SI{10}{\micro\hertz}, amplitude accuracy is better than \SI{1}{\milli\volt}. A stable quartz oscillator is embedded as frequency reference for deployable operation.

The acquisition system is mainly designed for the demodulation of phase signals from coherent optical interferometry, hence for measurements of the optical carrier frequency. Its characterization confirmed that the board performances are suitable for the analysis of optical frequency signals. Taking advantage of the optical-to-rf leveraging, the relative impact of the reported frequency measurement errors are of the order of \SI{e-19}, which is appropriate for many applications that do not require optical-clock-level accuracy. Notably, the operation principle provides this platform with all the typical advantages of a phase analyzer over, e.g., frequency counters. These include, among others, high sampling rates, intrinsically dead-time-free measurements, and well-defined user-selectable bandwidth. Moreover the phase measurement is combined with amplitude demodulation as in lock-in amplifiers, allowing to monitor the dynamic evolution of the carrier strength and synchronization capabilities.

An in-field demonstration of remote fiber sensing using distant ultrastable laser sources has been performed. The optical phase was acquired as carrier frequency fluctuations exploiting the self-heterodyne and heterodyne interferometric schemes, allowing to measure the relative noise of the laser sources and the fiber link noise.

Compared to commercial benchtop systems or \glspl{fpga}, the presented system is characterized by higher flexibility, stand-alone operation, simpler architecture, compact footprint, and much lower costs. It allows for real-time and continuous monitoring applications, supporting wide ranges of amplitude and frequency of the input signal. The embedded triggering mechanism permits to synchronize distant boards at the level of a few \si{\micro\second} without requiring additional channels or protocols, and allowing quantitative signal comparisons and correlation analysis. The \gls{dsp} algorithm optimization enables high performances also on a cheap and reliable \gls{mcu}. The platform represents an efficient possibility in deployable long-term experiments. Particularly interesting applications are related to fiber noise sensing, such as for earthquake monitoring, or metrological comparisons of distant laser sources using optical interferometry.

\section*{Acknowledgment}

\addcontentsline{toc}{section}{Acknowledgment}
The authors thank Open Fiber, in particular Francesco Carpentieri, and Metallurgica Bresciana for their support in providing and using the fiber testbed in the framework of Project MEGLIO. This work is supported by the European Metrology Program for Innovation and Research (EMPIR) Project 20FUN08 Nextlasers, which has received funding from the EMPIR program co-financed by the Participating States and from the European Union's Horizon 2020 research and innovation program.

\bibliographystyle{IEEEtran}
\bibliography{bibliography}

\vfill

\end{document}